\input{epsf.tex}
\documentclass[12pt]{iopart}
\begin{document}

\title{The Density Matrix Renormalization Group for finite Fermi systems}

\author{Jorge Dukelsky\dag\ and Stuart Pittel\ddag
\footnote[3]{To whom correspondence should be addressed
(pittel@bartol.udel.edu)} }

\address{\dag\ Instituto de Estructura de la Materia, Consejo Superior de\\
Investigaciones Cientificas, Serrano 123, 28006 Madrid, Spain}

\address{\ddag\ Bartol Research Institute, University of Delaware, Newark,\\
Delaware 19716, USA}

\begin{abstract}
The Density Matrix Renormalization Group (DMRG) was introduced by
Steven White in 1992 as a method for accurately describing the
properties of one-dimensional quantum lattices. The method, as
originally introduced, was based on the iterative inclusion of
sites on a real-space lattice. Based on its enormous success in
that domain, it was subsequently proposed that the DMRG could be
modified for use on finite Fermi systems, through the replacement
of real-space lattice sites by an appropriately ordered set of
single-particle levels. Since then, there has been an enormous
amount of work on the subject, ranging from efforts to clarify the
optimal means of implementing the algorithm to extensive
applications in a variety of fields. In this article, we review
these recent developments. Following a description of the
real-space DMRG method, we discuss the key steps that were
undertaken to modify it for use on finite Fermi systems and then
describe its applications to Quantum Chemistry, ultrasmall
superconducting grains, finite nuclei and two-dimensional electron
systems. We also describe a recent development which permits
symmetries to be taken into account consistently throughout the
DMRG algorithm. We close with an outlook for future applications
of the method.
\end{abstract}

\submitto{\RPP} 

\maketitle

\section{Introduction}

The properties of non-relativistic many-body quantum systems, as
arise in many branches of physics, are described by the
Schr\"{o}dinger equation. Unfortunately, the Schr\"{o}dinger
equation can be solved exactly in but a few exceptional cases,
either involving one-dimensional or symmetry-restricted
Hamiltonians or for systems with very few particles. For almost
all problems, reliable approximate solutions must be sought. Many
methods of approximately solving the Schr\"{o}dinger equation for
many-body quantum systems have been developed, often tailored to
the specific system or problem under investigation. Especially
challenging has been to come up with methods that can treat large
numbers of particles with great accuracy, particularly in the
vicinity of a phase transition. A major breakthrough was the
development of the numerical renormalization group (RG) method by
Wilson \cite{W0} in the mid-1980s, which provided for the first
time an efficient strategy for solving the Kondo problem
\cite{Kondo}. The enormous power of the RG method immediately
suggested its application to quantum lattices. It soon became
clear, however, that the Wilson RG method ran into difficulties
when dealing with such systems.

The Density Matrix Renormalization Group (DMRG) was introduced by
White in 1992 \cite{W1,W2,W3} in an effort to overcome the
limitations of Wilson's RG in describing one-dimensional (1D)
quantum lattice models. The new method was soon shown to be
extremely powerful, producing results for the ground state energy
of the S=1 Heisenberg chain that were accurate to twelve
significant figures \cite{W2}, well beyond the precision of
large-scale diagonalization methods combined with finite size
corrections or of Monte Carlo techniques.

The DMRG is an approximate variational procedure that is rooted in
Wilson's {\it onion picture} \cite{W4}. In the context of quantum
lattices, the idea is to start with a set of lattice sites and
then to iteratively add to it subsequent sites until all have been
treated. At each iteration, the dimension of the enlarged space
increases as the product of the dimension of the initial subspace
and that of the added site. The renormalization group procedure
consists of reducing the enlarged space to the same dimension as
the initial subspace and then transforming all operators to this
new truncated basis (renormalization).

One of the key features that distinguishes the DMRG from Wilson's
RG is the criterion by which the truncation is implemented. While
the Wilson RG retains the lowest Hamiltonian eigenstates of the
enlarged space, the DMRG uses a very different strategy. Here the
idea is to construct the reduced density matrix for the enlarged
space in the presence of a medium that approximates the rest of
the Hilbert space, then to diagonalize this density matrix, and
finally to maintain only those states with the largest density
matrix eigenvalues. This method of truncation is guaranteed to be
optimal in the sense that it maximizes the overlap of the
approximate (truncated) wave function with the wave function prior
to truncation.

Following the ideas just described, one can treat the entire set
of ``onion" layers. Depending on the manner in which correlations
between layers fall off and in the choice of the order in which
layers are treated, this can sometimes lead to an accurate
representation of the ground state of the system. Usually, it does
not, however, since the early layers know nothing of the physics
of those treated subsequently. This suggests the use of a
``sweeping" algorithm, whereby once all layers have been sampled,
we simply reverse direction and update them based on the
information of the previous ``sweep". Such a sweeping algorithm
can be iteratively implemented until acceptable convergence in the
results has been achieved.

The real-space DMRG (rDMRG), as described above, has been applied
extensively and with enormous success to many 1D systems
\cite{W3,WH}, such as spin chains and ladders \cite{SCL,zigzag},
t-J \cite{tJ} and Hubbard models \cite{Hub1}, models with disorder
\cite{Dis}, models with impurities \cite{Imp}, polymers
\cite{Pol}, and the Kondo model \cite{1DKondo}. Extension of the
method to 2D lattices \cite{2D} requires establishing a path that
covers all of the sites of the lattice. Since there are many
possible paths through a 2D lattice, it is critical that the path
chosen conform to the dominant correlations of the problem.
Electrons in small square or rectangular lattices with t-J or
Hubbard Hamiltonians \cite{2De} have been treated using this DMRG
strategy, though with much less precision than for 1D systems.
Problems in 3D are beyond the scope of the DMRG in real space.

A pedagogical introduction to the real space DMRG method together
with a summary of some of its most important applications can be
found in the proceedings of the workshop ``Density-Matrix
Renormalization" \cite{book} held in Dresden in September 1998.

Since then, there have been many important developments and
advances of the DMRG, which has opened up the possibility of its
application to finite Fermi systems. These new developments
originated with the introduction of a transformation of the DMRG
from real space to momentum space (kDMRG). The kDMRG method was
first discussed by Xiang \cite{xiang} and applied to the Hubbard
model in 2D. The results were not very encouraging, however, and
the method was therefore not pursued at that time for other 2D
quantum lattices.

This method was recovered in a slightly modified form in 1999 in
an effort to build a DMRG methodology appropriate to problems in
quantum chemistry \cite{W5,W6}. In this approach, the momentum
basis was replaced by a basis of single-particle orbitals, so as
to better accommodate the dynamics of the problem. The first
applications were reported in refs. \cite{W5,W6}. Several other
applications to problems in quantum chemistry were subsequently
reported \cite{q1,q2}.

A somewhat different extension of the DMRG methodology was
proposed in refs. \cite{duke1,duke2}, in the context of a study of
ultrasmall superconducting grains. In that work, a modification of
the algorithm -- called the particle-hole DMRG (phDMRG) -- was
proposed in order to optimally account for the crucial
correlations close to the Fermi energy. The method was shown to
work extremely well for superconducting grains up to very large
dimensions. The phDMRG was then introduced for possible
application in large-scale nuclear shell-model calculations
\cite{duke3,duke4,duke5}.

Several other important applications of the extended methodology
have also been reported recently, including applications  to
electrons in the lowest three Landau levels
\cite{shi1,shi2,shi3,shi4,shi5}.

The DMRG methodology, both in real space and in the versions
appropriate to finite Fermi systems, usually violates some
symmetries of the underlying Hamiltonian. While this is not
critical when studying certain features of the associated quantum
systems, it may be critical when dealing with others. There have
been some recent efforts to modify the DMRG algorithm
\cite{IRF1,IRF2,MG1,MG2} to permit the preservation of symmetries
throughout the iterative truncation process.

In this review, we summarize the rapid progress now being made on
the DMRG methodology and the wide variety of important
applications that have been reported recently for strongly
correlated finite Fermi systems. We begin in Section II with an
historical overview of the method and review its application in
real space to quantum lattices. In the same section, we describe
the basic features of the kDMRG extension and discuss its
application to quantum lattices in more than one dimension. In
Section III, we begin our focus on finite Fermi systems,
discussing the modifications to the general DMRG algorithm needed
in such cases and reviewing the applications to date. Then in
Section IV we discuss the issue of symmetry restoration,
summarizing recent developments and describing in some detail how
this new machinery can be used in the context of angular momentum
restoration. In Section V, we close the presentation by providing
some perspectives for future applications of the DMRG approach in
finite Fermi systems.

\section{Historical overview of the DMRG method}

\subsection{The many-body quantum problem}

A quantum many-body problem is completely defined by giving a
basis of one-body states, a Hamiltonian operator and the number of
particles. Usually, the one-body states are obtained from a
self-consistent mean-field calculation, such as Hartree-Fock (HF),
or by some optimized orbital basis. In atomic and molecular
physics \cite{QuantumChemistry}, for example,  the electrons move
in the attractive Coulomb field generated by the atomic nucleus.
The residual electron-electron interaction can be taken into
account in mean field to give a set single-electron energy states,
which then can serve as the starting basis. In nuclear physics
\cite{Nuclear}, the nucleon-nucleon mean field generates a set of
self-consistent single-nucleon levels, which can provide the
basis. Alternatively, we can ignore detailed self-consistency and
choose a basis for convenience. Convenient single-nucleon bases
used frequently in nuclear physics include the eigenstates of a
harmonic oscillator potential (all localized) and the eigenstates
of a Woods-Saxon potential (both localized and continuum states).

Though the main goal of this review is to present the latest
developments using the DMRG method in such finite Fermi systems,
we will begin with an historical overview, starting with the
development of the method for quantum lattices and its
applications in that domain. Here, the single-particle basis state
is replaced by a single active electron level, located at a given
site in real space.

One of the most well-known models of strongly interacting
electrons on a lattice is the Hubbard model \cite{Hub}. It
describes the hopping of electrons between nearest neighbor sites
(atoms) subject to a repulsive Coulomb interaction between them
when they occupy the same site. The Hamiltonian of the Hubbard
model takes the form

\begin{equation}
H=-t\sum_{<ij>\sigma }\left( c_{i\sigma }^{\dagger }c_{j\sigma
}+c_{j\sigma }^{\dagger }c_{i\sigma }\right)
+U\sum_{i}n_{i\uparrow }n_{i\downarrow } ~. \label{Hub}
\end{equation}
The operator $c^{\dagger}_{i \sigma}~(c_{i\sigma})$ creates
(annihilates) an electron on site $i$ with spin projection
$\sigma$. The first term represents the hopping of electrons with
intensity $t$.  The strength of the on-site Coulomb repulsion is
$U$ and $n_{i \sigma}=c^{\dagger}_{i \sigma} c_{i \sigma}$ denotes
the number operator for an electron on site $i$ with spin
projection $\sigma $.

Assuming that we have a system with $N$ sites, the exact solution
of the problem amounts to constructing the many-body basis of
states in the appropriate Hilbert space, calculating the matrix
elements of the Hamiltonian in this basis and then diagonalizing.
In the Hubbard model, each site admits precisely four states, as
it can be empty, it can contain an electron with spin up, it can
contain an electron with spin down, or it can be fully occupied by
two electrons.  Thus, the dimension of the Fock space is naively
$4^{N}$, growing exponentially with the number of sites. Because
of conservation of symmetries in the many-body basis, {\em e.g.}
total particle number, total spin, total momentum, etc., the
Hamiltonian matrix separates into blocks, each with reduced
dimension. Nevertheless, the order of magnitude of each block
dimension is still comparable to the full Fock space dimension and
thus likewise grows exponentially with the number of lattice
sites. For this reason, exact large-scale diagonalization using
the Lanczos or Davidson algorithms is only feasible for a small
number of sites. The largest reported exact diagonalization of the
Hubbard model, for example, was for a 4$\times$4 2D lattice.

Exact diagonalization, combined with finite size extrapolation,
has proven very useful when trying to understand the gross
features of several lattice models. Nevertheless, we would like to
do better, by being able to treat larger systems. Obviously, if we
want to treat larger systems we have to implement an efficient
truncation strategy, one that permits an accurate description of
the low-energy properties of the system while only keeping a
reasonably small part of the full Hilbert space.

In the finite Fermi systems that arise in Quantum Chemistry (QC)
and in Nuclear Structure (NS), similar issues arise as regards the
size of the Hilbert space and the need for an efficient truncation
strategy. In those fields, a variety of methods have been
developed and used with great success in the study of small
systems, such as simple molecules and light nuclei. In QC, for
example, the Configuration Interaction (CI) method
\cite{QuantumChemistry} working in a HF basis with a limited set
of particle-hole states can be used to build a truncated many-body
Hilbert space in which to diagonalize the Hamiltonian and obtain
approximate solutions of the Schr\"{o}dinger equation. In NS, the
Shell Model (SM) \cite{ShellModel} starts with a nuclear mean
field generated by the average interaction of one nucleon with all
the others. For simplicity, the mean field is often modelled by a
harmonic oscillator potential with a spin-orbit coupling, defining
a set of shells well separated in energy. The nuclear interaction
expressed in this basis is typically diagonalized for a given
nucleus within one major shell for each type of nucleon. Despite
the enormous truncation inherent in both the CI approach in QC and
the SM approach in NS, the methods are limited as to the systems
that can be treated accurately. In NS, for example, {\em exact}
shell-model diagonalization techniques cannot be applied to nuclei
beyond the $1f-2p$ shell, leaving outside the scope of detailed
shell-model microscopy most medium-mass and heavy nuclei.

Going beyond the limits of exact diagonalization techniques, even
within the restricted bases just described, requires a new kind of
truncation strategy. One that has been used with great success in
many systems is the Monte Carlo approach. Another is the
renormalization group (RG) approach \cite{W0}, the subject of this
review. In this method, a set of renormalization transformations
that gradually grow the size of the system while keeping the
dimension of the configuration space constant is introduced. The
procedure is implemented iteratively. At each step, the system is
grown by the addition of new degrees of freedom. The enlarged
system is then truncated to the same size as before, and a
renormalization is implemented which transforms the Hamiltonian
and all other relevant operators to the truncated space. The
numerical renormalization group of Wilson represented a
breakthrough in the application of RG ideas to a non-trivial
strongly correlated system, the single-impurity Kondo problem
\cite{Kondo}. Later on it became clear that the Wilson RG (WRG)
could not be applied to other strongly correlated lattice problems
\cite{Chui}, for reasons that we will discuss below. Nevertheless,
the method and its failure provide the basis for the development
of the DMRG.

\subsection{The Wilson Renormalization Group}

In this subsection, we briefly describe the numerical
renormalization group of Wilson for a generic one-dimensional
lattice. While we will have in mind for concreteness the Hubbard
Hamiltonian (\ref{Hub}), the formalism we will describe is general
enough to incorporate  other one-dimensional lattice models,
including the Kondo problem. In fact, Wilson first formulated his
numerical RG after mapping the Kondo problem onto a 1D lattice in
energy space.

\begin{figure}
\begin{center}
\hspace{2.5cm} \epsfxsize=4.5cm \epsfysize=4cm\epsfbox{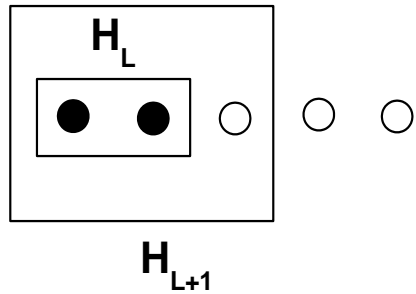}

\caption{\label{WRGsites}Schematic illustration of the sites on a
one-dimensional lattice and the Wilson Renormalization Group
method of enlarging blocks of sites. The sites denoted by a bullet
$\bullet$ have already been treated, and the next site to the
right is in the process of being added.}
\end{center}
\end{figure}

The starting point of a WRG calculation (and as we will discuss
later the DMRG as well) is a small system for which we know all of
the states and for which we can calculate the matrix elements of
all relevant operators including the Hamiltonian. We begin,
therefore, by considering a set of $L$ contiguous lattice sites -
which we will call a {\em block} - with $L$ small enough so that
we can exactly diagonalize the Hamiltonian in the associated
subspace. Figure \ref{WRGsites} illustrates a chain of lattice
sites, with the filled dots representing the sites of the initial
system - $L=2$ in this case - and the open dots representing all
other sites in the problem. For $L=2$, the number of states in the
block for the Hubbard model is $m=16$.

The WRG procedure begins with a calculation of the matrix elements
of all relevant operators in the initial block. The relevant
operators are those that enter the Hamiltonian, namely $c_{i\sigma
}^{\dagger }$, $c_{i\sigma }$, $n_{i\sigma }$, and the Hamiltonian
(\ref{Hub}) itself.

Then we continue as follows:

${\bf 1.}$ Add the site $L+1$ and calculate the matrix elements of
all operators in the enlarged block space of $4m$ states. [Note:
If we wish, we can partition the states of a site and add only 2
states, rather than 4 in each step. When this is done, the
enlarged block space has $2m$ states instead of $4m$.]

${\bf 2.}$ Diagonalize the Hamiltonian and select the lowest $m$
eigenvalues and the corresponding eigenvectors.

${\bf 3.}$ Transform all operators in the enlarged block to a
truncated basis involving these $m$ states only.

${\bf 4.}$ If $L+1$ is lower that the total number of sites $N$,
substitute $L+1$ for $L$ and repeat steps ${\bf 1 \rightarrow 3}$.

The key ingredient in the Wilson renormalization strategy is step
${\bf 2}$, whereby $m$ states are selected from the $4m$ (or $2m$)
states of the enlarged block {\em by a purely energetic
criterion}. Clearly, the accuracy of the method relies on the
assumption that the low-energy properties of the system are fully
contained in the low-energy states of the individual blocks into
which the system was partitioned. Indeed, this was the case for
the Kondo problem, which was solved by Wilson by transforming it
to a discretized energy basis, and for other impurity models like
the Anderson model \cite{Anderson}.

The Kondo model describes the antiferromagnetic interaction of the
conduction electrons with a localized impurity. Its Hamiltonian in
a discretized energy basis takes the form

\begin{equation}
H=\frac{1}{2}\left( 1+\Lambda ^{-1}\right) \sum_{n=0}^{\infty
}\Lambda ^{-n/2}\left( f_{n\mu }^{\dagger }f_{n+1\mu }+f_{n+1\mu
}^{\dagger }f_{n\mu }\right) +2J\rho f_{0\mu }^{\dagger }\sigma
_{\mu \nu }f_{0\nu } ~,\label{Kondo}
\end{equation}
where $f_{n\mu }^{\dagger }\left( f_{n\mu }\right) $ creates
(annihilates) an electron in the energy interval $n$ with spin
$\mu $, $J$ is the strength of the antiferromagnetic interaction
between the conduction electrons and the impurity, $\rho $ is the
density of states and $\Lambda $ is a discretization parameter
defining the intervals of energy. For $\Lambda >1$ and beginning
from the origin ($n=0$) of the linear chain, the interaction along
the chain decays exponentially, assuring the validity of the
truncation procedure. The WRG was the first numerical
implementation of the RG to a non-perturbative problem like the
Kondo model, for which it had enormous success. As such, it is
still used to this day to study problems with one or few
impurities.

Unfortunately, the Wilson RG cannot be applied to other lattice
problems. For example, when it is applied to 1D Hubbard models it
begins to deviate significantly from the exact results in the
fourth iteration. Obviously, the problem resides in the fact that
the truncation strategy was based solely on energy arguments. The
solution to this problem was proposed by White with his
introduction of the DMRG, which we turn to in the next subsection.

\subsection{\protect\bigskip The Density Matrix Truncation Strategy}

\begin{figure}
\begin{center}
\hspace{1.5cm} \epsfxsize=4.2cm \epsfysize=2cm\epsfbox{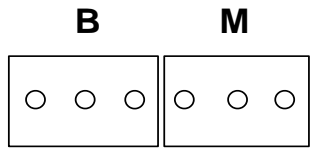}
\end{center}
\caption{\label{DMRGBlocks1}Schematic illustration of the
two-block structure of the Density Matrix Renormalization Group. }
\end{figure}

The essential limitation of the WRG, as discussed in the previous
subsection, is the method used for truncation at each RG iteration
step. Once the enlarged block is constructed by adding the site
$L+1$ in step $1$, the Hamiltonian in the $L+1$ -- site subspace
of the total Hilbert space is diagonalized, without considering
the coupling to the rest of the space. Obviously, the method will
work well when this coupling is perturbatively small. This is the
case for Hamiltonians  that can be transformed to the form
(\ref{Kondo}).  In general, however, the WRG performs very poorly
for strongly correlated lattice systems. In searching for a more
suitable method of truncation, White analyzed the case of a block
immersed in a medium representing the rest of the system, as
depicted in figure \ref{DMRGBlocks1}.

Assume that we know the ground state $\left| \Psi \right\rangle $
of the system and the complete set of $p$ states of the block $B$
and $t$ states of the medium $M$. Then the ground state of the
full system $\left| \Psi \right\rangle $ can be expanded in the
product space of the block and the medium as

\begin{equation}
\left| \Psi \right\rangle =\sum_{i=1,p}\sum_{j=1,t}\Psi
_{ij}\left| i\right\rangle _{B}\left| j\right\rangle _{M} ~.
\label{Psi}
\end{equation}

Our aim is to determine the best approximation $\left| \Psi
^{\prime }\right\rangle $ to the ground state $\left| \Psi
\right\rangle $ when we select only $m$ states of the block, with
$m<p$. To show how this is done, we will make use of the concept
of the singular value decomposition (SVD) of a matrix \cite{svd}.

The singular value decomposition of the wave function matrix
$\Psi_{ij}$ in (\ref{Psi}) permits its factorization as

\begin{equation}
\Psi _{ij}=\sum_{\alpha =1,p}a_{\alpha }u_{i\alpha }v_{\alpha j}
~, \label{svd}
\end{equation}
where the $a_{\alpha }$ are real positive numbers greater than or
equal to 0, $u$ is an orthogonal square matrix of dimension $p$
and $v$ is a column orthogonal rectangular matrix of dimension
$p\times t$. The coefficients $a_{\alpha }$ are known as the
weight factors of the SVD.

Up to this point, everything is exact. Now consider what happens
if some of the weight factors in (\ref{svd}) are zero or very
small. If that is the case, then $\Psi $ can be very well
approximated by discarding them, {\em i.e.} by truncating the sum
in (\ref{svd}) to those with the largest weights. If fact, it can
be shown that the approximate wave function factorized as in
(\ref{svd}) but retaining the largest $m$ weight factors has the
largest overlap with the exact wave function for any factorization
involving $m$ terms.

In what follows, we will use the singular value decomposition as a
convenient factorization for deriving the conditions for optimal
truncation in a density matrix formalism. It is interesting to
note, however, that the same approach has been used to develop a
powerful new technique for numerically diagonalizing the nuclear
shell-model problem without actually introducing the DMRG
\cite{papen}.

Now consider the reduced density matrix of the block associated
with the ground state wave function (\ref{Psi}), which is defined
as

\begin{equation}
\rho^B _{ij}=\sum_{k=1,t}\Psi _{ik}\Psi _{jk}  ~,\label{ro}
\end{equation}
with the summation running over the $t$ states of the medium.
Inserting the factorization (\ref{svd}) into (\ref{ro}) and using
the orthogonality relations of the matrices $u$ and $v$ and the
normalization of the wave function, we can express the reduced
density matrix (\ref{ro}) as

\begin{equation}
\rho^B _{ij}=\sum_{\alpha =1,p}a_{\alpha }^{2}u_{i\alpha
}u_{\alpha j} ~.\label{rop}
\end{equation}

From this we see that $u$ is the matrix of eigenvectors of the
reduced density matrix and $\omega _{\alpha }=a_{\alpha }^{2}$ are
the corresponding eigenvalues. Each eigenvalue $\omega _{\alpha }$
represents the probability of finding the block state $u_{\alpha}$
in the ground state $\left| \Psi \right\rangle $ of the coupled
system. The normalization of the wave function $\Psi $ guarantees
that the trace of $\rho^B $ is 1. Clearly a truncation to $m$
states implies the selection of the largest $m$ eigenvalues and
corresponding eigenvectors. The discarded $p-m$ states contribute
with a probability $1-P_{m}=1-\sum_{\alpha =m+1}^{p}\omega_{\alpha
}$. This parameter provides an important control on the accuracy
of the truncation.

Up to now, we have considered the system as being in a pure state.
A similar analysis can be carried out for systems in a mixed
state. Mixed states naturally arise when considering a system at
finite temperature ($T$).  They can even arise at $T=0$ if we
decide to target several states with the DMRG truncation strategy.
By using mixed density matrices associated with several states, we
can build a truncation algorithm whereby the gradually enlarging
block optimally reflects the physics of all of the associated
states and not just the physics of the ground state.  The most
important states of the block, when it is immersed in a
mixed-state system, are those corresponding to the largest
eigenvalues of the mixed density matrix,

\begin{equation}
\rho ^B_{ij}=\sum_{a}W_{a}\sum_{k=1,t}\Psi _{ik}^{a}\Psi _{jk}^{a}
~, \label{romix}
\end{equation}
where $W_{a}$ is the probability with which a given pure state
$\left| \Psi ^{a}\right\rangle $ enters in the mixed state. In the
case of systems at finite temperature, $W_{a}$ is the Boltzmann
factor. For a mixed state at T=0, we have the freedom to select
the weights so as to assign a relative importance to each state.

\subsection{\protect\bigskip The DMRG Method
for Quantum Lattices: Real space}

In the previous subsections we described how we can choose the
optimal states of a part of a quantum system from its density
matrix in the presence of a medium approximating the rest of the
system. Here, we show how this can be used in a systematic
treatment of many-site quantum lattices.

We begin by verbally detailing the key ingredients of the method.
The basic idea is to {\em systematically} take into account the
physics of {\em all}  sites on the lattice.  The procedure begins
by starting at the left (or right) edge of the lattice and then
gradually growing the size of the space that has been taking into
account by adding the other lattice sites.  At each step of this
iterative procedure, a truncation is implemented, so as to
optimally take into account the effect of the most important
states within that part of the problem. The calculation is carried
out as a function of the number of states that are maintained
after each iteration. This parameter, which we call $m$, is
gradually increased and the results are plotted against it.
Experience  suggests that the results converge exponentially with
$m$. Thus, once we achieve changes with increasing $m$ that are
acceptably small we terminate the calculation.

In the following, we show how this strategy is implemented at two
different levels of complexity, referred to as the infinite
algorithm and the finite algorithm, respectively. We then describe
some technical details of importance to the computational
feasibility of applying the method.

\subsubsection{The Infinite Algorithm.}

The infinite algorithm, as we will soon see, can only be applied
meaningfully in the presence of a uniform (or translationally
invariant) lattice. Thus, in what follows we assume that the
lattice under discussion satisfies this criterion.

\begin{figure}
\begin{center}
\hspace{2cm} \epsfxsize=5.5cm\epsfysize=6.9cm \epsfbox{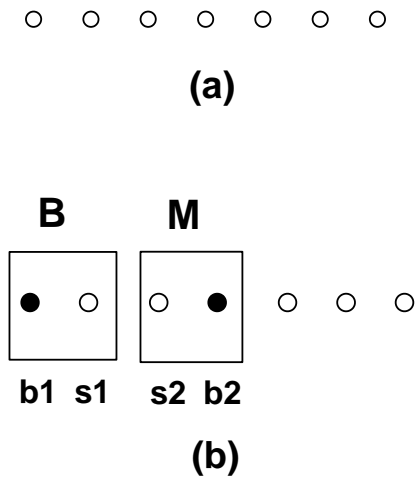}
\end{center}
\caption{Schematic illustration of the infinite DMRG algorithm.
(a) The sites on a uniform one-dimensional lattice that will be
used in subsequent discussion of the DMRG algorithm. (b) The
two-block structure of the algorithm. A bullet ($\bullet$) refers
to a block of sites and a circle ($\circ$) to a single site. $B$
refers to the enlarged block, which is comprised of $b1$ and $s1$.
$M$ refers to the enlarged medium, which is comprised of $s2$ and
$b2$.} \label{DMRG2}
\end{figure}

Consider such a lattice, as illustrated in figure \ref{DMRG2}a.
Assume that each site on the lattice admits $l$ states. In the
case of a spin-$1/2$ chain, $l=2$, corresponding to spin up or
spin down. In the case of the t-J model, $l=3$, as the site can
also be empty. Finally, as noted earlier, in the Hubbard model a
site can also have both spin-up and spin-down particles and $l=4$.

We now begin to add sites, by assumption from the left. Until we
have included enough sites $k$ so that $l^k > m$, no truncation is
required. We thus begin our discussion after we have added a
sufficient number of sites to be in the truncation regime. We
assume that we have already treated the leftmost $k$ lattice sites
and have identified the most important $m$ states.

At that point, we can view the problem as illustrated in figure
\ref{DMRG2}b. The collection of $k$ sites that have already been
treated is represented by a $\bullet$ and denoted $b1$.
Immediately to its right is another site, which contains $l$
states and is denoted $s1$. We then have another site to its
right, which also contains $l$ states and is denoted $s2$. After
that we have another block, representing the optimal states from
sites $k+3 \rightarrow 2k+2$. Because  of the translational
invariance of a uniform quantum lattice, all parts of the lattice
look the same. As a consequence, once we know the optimal states
from sites $1 \rightarrow k$ we also know the optimal states from
any other group of $k$ contiguous sites. This second block is
denoted $b2$. And then finally we have other sites to the right
that are still to be considered.

We then consider as an enlarged block the $m \times l$ states
obtained by adding the states of site $s1$ to those of block $b1$.
We will use the Density Matrix strategy described in the previous
subsection to truncate from these $m \times l$ states to the
optimal $m$ states, the same number that we had before we enlarged
the block. As the medium in which to implement the truncation, we
use the set of $m \times l$ states of the site $s2$ coupled to the
block $b2$. The rationale for growing both the block and the
medium at the same time is discussed in ref. \cite{W2}.

Following implementation of the density matrix truncation, we make
the following replacements:

\begin{itemize}

\item
The set of $m$ states of the enlarged block now becomes block
$b1$.

\item
The next available site to its right becomes $s1$

\item
The next site after that becomes $s2$.

\item
The truncated set of $m$ states from the next $k+1$ sites becomes
block $b2$.
\end{itemize}

This leaves us in the same position as in figure \ref{DMRG2}b,
with a block of $m$ states, followed by a site of $l$ states, then
another site of $l$ states and finally a block of $m$ states. As
before, we build enlarged blocks out of the first two and out of
the second two, respectively. We then truncate the enlarged block
from $b1$ and $s1$ to the same $m$ states as before. And we then
use the translational symmetry of the problem to form another set
of four objects, $b1$, $s1$, $s2$, $b2$, and continue the
iteration. This iterative process can be continued until the block
includes half of the sites. At this point, the superblock contains
the physics of all of the sites and thus gives a representation of
the entire problem.

Such a procedure, in which we grow the system from left to right
(or from right to left) until all sites have been treated is
called the infinite DMRG algorithm. Clearly it requires that the
lattice be uniform, so that we can use the translational
invariance of the problem to relate the collective structure of
the medium to that of the block. In such cases, the accuracy of
the procedure still depends on how rapidly the correlations
between sites fall off. {\em Exact} numerical results can be
obtained when the correlations are predominantly between fairly
close neighbors or, more precisely, when the correlation length is
finite. A typical example is the $S=1$ Heisenberg chain \cite{W3}.

\subsubsection{The Finite Algorithm and Sweeping.}

Often the above criteria for success of the infinite algorithm are
not satisfied and it is necessary to do better. When the lattice
is not unform, there is no natural way to build a medium that
properly reflects correlations in that part of the lattice. When
the correlations do not fall off significantly rapidly, there is
no way to take into account the effects of distant sites - those
that are not part of the medium - in defining the collective
structure of the block.  A way to overcome these limitations is to
use the so-called finite algorithm, which we now describe.

The basic idea of the finite algorithm is that we {\em sweep}
forwards and backwards through the set of sites, at each stage
updating the information from the previous sweep. We will continue
to focus here on quantum lattices, and in that context show how
the method is implemented. We will first carry out the discussion
for a uniform lattice and then mention the modifications needed
when the lattice is not uniform.

As before we begin by growing the system from the left. For a
unform lattice, as depicted in fig. \ref{DMRG2}a, we follow the
infinite algorithm until half of the sites have been treated in
the left block and half in the right block. In each iteration we
store on hard disk all relevant matrix elements within the block
for later use. From this point on, the superblocks will involve
all the $N$ sites of the finite lattice. We then continue to grow
the sites in the left block, one at a time. For each enlarged
block, we use for the medium the remaining set of sites from the
right half. All information on that set of sites is obtained by a
reflection from the stored information of the left block from an
earlier iteration, assuming inversion symmetry around the center
of the lattice.

We continue this process until the left block has been grown to
include all sites but three. At this point, we have stored
information on all subgroups of sites, from 2 through $N-3$. That
information includes the matrix elements of all sub-operators of
the Hamiltonian, within the space of states being maintained.

If the lattice is not uniform, we must use a slightly different
strategy, since there is no simple prescription for relating the
states of the medium to those of the block. One possibility is to
use the Wilson RG procedure, growing the system one site at a time
and truncating by energy considerations \cite{xiang}.
Alternatively, one can use the infinite algorithm, beginning with
blocks at the two ends and gradually expanding them by adding
sites in between in the direction of the lattice center
\cite{1DKondo}. In either case, we continue until all sites have
been treated and information on all subgroups has been calculated
and stored.

\begin{figure}
\begin{center}
\hspace{2cm} \epsfxsize=6.8cm \epsfysize=4.0cm \epsfbox{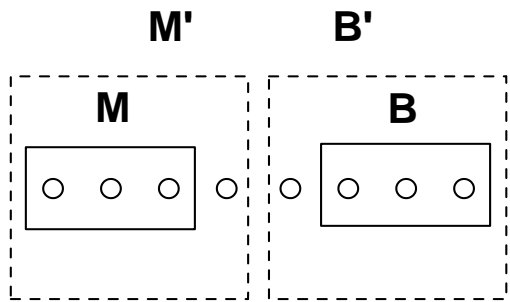}
\end{center}
\caption{Schematic illustration of the finite DMRG algorithm. {\bf
B} denoted the initial block and {\bf B$^{\prime}$} the enlarged
block; {\bf M} denotes the initial medium and {\bf M$^{\prime}$}
the enlarged medium. } \label{DMRG3}
\end{figure}

At this point, whether dealing with uniform or non-uniform
lattices, we reverse the process and {\em sweep} from the right,
as illustrated in figure \ref{DMRG3}. We begin to enlarge the
right block, by adding one site at a time. The medium for this
enlarged block is the set of sites on the left, which consists of
the block $M$ and the extra site. Note that all relevant
information on the block $M$ was stored in the previous sweep.

At this point, we reverse again, sweeping from left to right. The
sweep process is continued iteratively until the results from one
sweep are acceptably close to those from the previous sweep. The
comparison can be done for the ground-state energy and for the
density matrix eigenvalues of the retained states.

Note that the first (or warm-up) stage, in which we sweep through
the sites for the first time, need not necessarily produce very
good results, since we still have the opportunity to improve on
them in subsequent iterations. It is for this reason that the
finite algorithm can provide a good approximation to the physics
of non-unform lattices.

\subsubsection{Some technical details.}

For the purposes of this discussion, we assume that the
Hamiltonian describing the system under investigation contains
one- and two-body terms only and can be written as

\begin{equation}
H=\sum_{m}\varepsilon _{m}  a _{k}^{\dagger }a _{k} ~+~
\frac{1}{4} \sum_{mnop} V_{mnop}~a _{m}^{\dagger }a _{n}^{\dagger
} a_{p}a _{o} ~.
\end{equation}

The above Hamiltonian is itself built up out of other operators,
which we referred to earlier and will continue to refer to as the
Hamiltonian sub-operators. They are

\begin{equation}
a^{\dagger}_m ~,~ a^{\dagger}_m a_n ~,~ a^{\dagger}_m
a^{\dagger}_n~,~ a^{\dagger}_m a^{\dagger}_n a_p ~,~ a^{\dagger}_m
a^{\dagger}_n a_p a_o ~, \label{sub-ops}
\end{equation}
plus their hermitean adjoints.

A key to the computational feasibility of the DMRG algorithms
described earlier in this subsection is that it can be readily
automated, using information from previous steps. Here we discuss
this point in somewhat more detail.

\begin{figure}
\begin{center}
\hspace{2cm} \epsfxsize=4.2cm \epsfysize=2.5cm \epsfbox{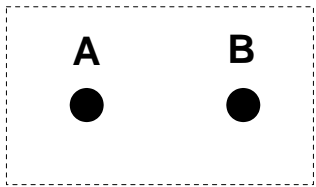}
\end{center}
\caption{Schematic illustration of block enlargement in the DMRG.
{\bf A} denotes the initial block and {\bf B} denotes the block
(or site) that is going to be added to it. } \label{DMRG4}
\end{figure}

As we have seen, the DMRG approach involves a systematic
enlargement of blocks, followed by a truncation to the same block
size as before the enlargement. Consider a representative block
enlargement as illustrated in figure \ref{DMRG4}.  The original
block $A$ with dimension $d_A$ was obtained from an earlier step,
at which point the matrix elements of all sub-operators of the
Hamiltonian (see Eq. (\ref{sub-ops})) were calculated and stored.
We add to it another block $B$ of dimension $d_B$, which can
either be a block obtained in an earlier step or a single lattice
site. If it is a block obtained from an earlier step, the matrix
elements of all sub-operators of the Hamiltonian were calculated
and stored at that time for the corresponding set of states. If it
is a single lattice site, the matrix elements of the Hamiltonian
sub-operators were stored at the outset of the calculation. In
either case, the matrix elements of all Hamiltonian sub-operators
are accessible in storage when the enlargement of the block is
implemented.

At this point, we wish to calculate the matrix elements of the
Hamiltonian sub-operators in the enlarged block, {\em i.e.} in a
space of product states

\begin{equation}
|i,j \rangle = |i \rangle_A  ~|j \rangle_B ~,
\end{equation}
where $i=1,...,d_A$ and $j=1,...,d_B$.

Consider as a specific example the matrix elements of the one-body
operator $a^{\dagger}_{m} a_{n}$ in this basis. They can be
obtained as a sum over terms involving products of matrix elements
in space $A$ and space $B$, {\em viz:}

\begin{eqnarray}
\langle i, j | a^{\dagger}_{m} a_{n} | k,l \rangle && =_A\langle i
| a^{\dagger}_{m} a_{n}
| k \rangle_A ~\delta_{j,l}~\nonumber \\
 &&+~
_B\langle j | a^{\dagger}_{m} a_{n} | l \rangle_B ~\delta_{i,k} \nonumber \\
&&+ (-)^{n_{k_A}} ~_A\langle i | a^{\dagger}_{m} | k \rangle_A ~
_B\langle j | a_{n} | l \rangle_B \nonumber \\
&&- (-)^{n_{k_A}}~ _A\langle i | a_{n} | k \rangle_A ~ _B\langle
j | a^{\dagger}_{m} | l \rangle_B ~, \nonumber \\
\label{enlarge}
\end{eqnarray}
where $n_{k_A}$ is the number of particles in state $|k
\rangle_A$. Depending on whether the operator indices refer to the
sites in block $A$ or to those in block $B$, a different term in
the sum applies. For example, if both $m$ and $n$ refer to sites
in block $A$, it is the first term that applies. Since all of the
matrix elements in Eq. (\ref{enlarge}) were calculated and stored
earlier, it is straightforward to compute the matrix elements in
the enlarged block.

While we have focused for specificity on the one-body
particle-hole operator $a^{\dag}_{m} a_{n}$, it should be clear
that the same remarks apply to all of the operators in Eq.
(\ref{sub-ops}). Obviously, the number of terms in the sum
increases with the number of creation and/or annihilation
operators that define the composite operator being treated.

Another key step of the DMRG algorithm is renormalization, whereby
all sub-operators of the Hamiltonian are transformed to the
truncated basis of block states emerging from the density matrix
diagonalization, {\em viz:}

\begin{equation}
\rho^B |\alpha>  = \omega^B_{\alpha} | \alpha >  ~,
\end{equation}
where
\begin{equation}
|\alpha>=\sum_{i}X^{\alpha}_i |i>~.
\end{equation}

The matrix elements of any Hamiltonian sub-operator $O$ in the
$4m$-dimensional enlarged block space $|i>$ can be transformed to
the truncated $m$-dimensional block space of states $|\alpha>$
according to

\begin{equation}
O_{\alpha\beta} = \sum_{ij}{X^{\alpha}_i}^* X^{\beta}_j O_{ij} ~,
\end{equation}
and then stored for use in a subsequent step of the algorithm.

\subsection{\protect\bigskip The DMRG Method for
Quantum Lattices: Momentum space}

In the previous two subsections, we reviewed the key steps in
implementing the DMRG algorithm for quantum lattices. The method
described worked directly in the real space of lattices, growing
the system through the addition of sites.

A key to the success of the DMRG algorithm is being able to define
an appropriate order in which to add layers of the onion. In the
case of lattices, this means an appropriate order in which to add
sites. In one dimension, there is a natural order, namely to add
sites either from the right or from the left, always adding the
next available ``nearest neighbor"  site(s). This is natural since
the interactions that couple sites tend to fall off rapidly with
increasing distance between them. But what about lattices in more
than one dimension? As an example, consider a 2D square lattice.
Near any given site there are four ``nearest neighbor" sites. What
is the correct path through this set of lattice sites to consider?

For such reasons, it may be useful when dealing with lattices in
more than one dimension to consider an alternative labelling of
the states of the system, one that is more conducive to a natural
ordering.

In 1994, Xiang \cite{xiang} proposed the possibility of carrying
out DMRG calculations in momentum space, rather than in real
space. In this section, we review the new ideas that emerged from
this proposal. As we will see in the next section, this idea
provides the inspiration for the algorithms that have since been
developed for finite Fermi systems, the principal focus of this
review.

To make the ideas most transparent, we will carry out the
discussion in terms of the two-dimensional Hubbard model with
nearest-neighbor hopping, as was done in the original paper of
Xiang. The Hamiltonian for this model was given earlier in Eq.
(\ref{Hub}).

The operator that creates a particle in Bloch state $\bf{k}\sigma$
with spin $\sigma$ and momentum $\bf{k}$ and the one that creates
a particle with the same spin $\sigma$ but at lattice site $i$ are
related according to

\begin{equation}
c^{\dagger}_{\bf{k}\sigma} = \frac{1}{\sqrt{N}} \sum_i e^{i\bf{k}
\cdot \bf{r}_i } c^{\dagger}_{i\sigma} ~,
\end{equation}
where $N$ is the total number of sites in the lattice and
$\bf{r}_i$ is the coordinate-space location of site $i$.

Using this, it is straightforward to transform the Hubbard
Hamiltonian (\ref{Hub}) to momentum space,

\begin{equation}
H = \sum_{\bf{k}\sigma} \epsilon_{\bf{k}} \hat{n}_{\bf{k}\sigma}
~+~ \frac{U}{N} \sum_{\bf{p,k,q}} c^{\dagger}_{\bf{p-q}\uparrow}
c^{\dagger}_{\bf{k+q}\downarrow} c_{\bf{k}\downarrow}
c_{\bf{p}\uparrow} ~, \label{kHub}
\end{equation}
where
\begin{equation}
\hat{n}_{\bf{k}\sigma} = c^{\dagger}_{{\bf k}\sigma}c_{{\bf
k}\sigma}
\end{equation}
is the number operator for particles with spin $\sigma$ and
momentum $\bf k$ and
\begin{equation}
\epsilon_{\bf{k}}=-t\sum_{<i,j>} e^{i{\bf k\cdot (r_i-r_j)} }
\end{equation}
is the energy dispersion of the particles, which is explicitly
diagonal in the momentum basis.

In particular, by diagonalizing the one-body hopping term for a 2D
lattice we find that

\begin{eqnarray}
\epsilon_{\bf k} &=& -2t~(cos~k_x + cos~k_y) ~, \nonumber \\
{\bf k}&=&(2\pi n_x/L ~,~ 2\pi n_y/L)~, \nonumber  \\
n_x,~n_y &=& -L/2+1, ...,L/2 ~;~ L=\sqrt{N} ~.
\end{eqnarray}

In table \ref{Hubbard12}, we give the momentum ``spectrum" for a
2D 4$\times$4  lattice. Possible values for $k_x$ and $k_y$ in
this case are -$\pi$/2, 0, $\pi$/2, and $\pi$. What is clear is
that there is a large degeneracy at half filling. While it is
natural to consider filling momentum-space levels in order of
increasing energy dispersion, it is clear that there will be a
problem with degeneracies (or near degeneracies when more general
hopping terms are used). Thus, the rationale we gave above for
introducing momentum space - namely to provide a more natural
ordering of levels (onion layers) -  may not turn out to be
warranted. Nevertheless, we will discuss what happens when the
methodology is in fact used. Before doing so, however, we will
first discuss some issues related to the implementation of the
momentum-space DMRG algorithm, which we subsequently refer to as
the kDMRG.

\begin{table}
\caption{\label{Hubbard12}Energy dispersion for the 2D $4 \times
4$ Hubbard model.}
\begin{indented}
\item[]\begin{tabular}{ccc}
\br
$\epsilon/t$ & Degeneracy & ($k_x$,$k_y$)  \\
\mr
-4& 1 & (0 , 0)  \\
-2& 4 & (-$\pi$/2,0), (0,-$\pi$/2),  ($\pi$/2,0), (0,$\pi$/2)  \\
0 & 6 & (-$\pi$/2,-$\pi$/2), (-$\pi$/2,$\pi$/2), ($\pi$/2,-$\pi$/2),
($\pi$/2,$\pi$/2), ($\pi$,0), (0,$\pi$) \\
2 & 4 & (-$\pi$/2,$\pi$),  ($\pi$/2,$\pi$), ($\pi$,-$\pi$/2), ($\pi$,$\pi$/2)  \\
4& 1 & ($\pi$,$\pi$)\\ \br
\end{tabular}
\end{indented}
\end{table}

\subsubsection{The size of the problem.}

As emphasized by Xiang in his original paper, the two-body
interaction appearing in (\ref{kHub}) contains of order $N^3$
terms, since each of the sums goes over $N$ possible momentum
values. Thus, a straightforward application of the DMRG algorithm
in momentum space would require iteratively calculating and
storing matrix elements of $O(N^3)$ operators, which would
obviously be very time and memory consuming for a large number of
sites. Xiang figured out a very clever way to group the operators,
so as to dramatically reduce the number whose matrix elements need
to be calculated and stored. Since an analogous strategy will be
required in all extensions of the usual rDMRG method, we will now
discuss how this is accomplished.

Consider the following set of operators, operating in a subspace
$A$ of the full 2D Hubbard Hilbert space:

\begin{eqnarray}
 a_0(p,\sigma) =c_{p\sigma}  & ~~~~~~(p,\sigma)\in{A} \nonumber \\
a_1(p,\sigma) = \sum_{q} a_0^{\dagger}(q,\sigma) a_0(p+q,\sigma) &
\nonumber \\
a_2(p)=\sum_{q}a^{\dagger}_0(q,\uparrow)a_0(p+q,\downarrow)
& \nonumber \\
a_3(p,\sigma)=\sum_{q_1q_2} a^{\dagger}_0(q_1,-\sigma)
a_0(q_2,-\sigma)a_0(p+q_1-q_2,\sigma) & ~~~~~~(p,\sigma)\not\in{A}
\nonumber \\
a_4(p)=\sum_q a_0(q,\downarrow)a_0(p-q,\uparrow)~. & \nonumber \\
\end{eqnarray}
The total number of such composite operators is $6N$,
significantly less than the total number of sub-operators of the
full Hubbard Hamiltonian in momentum space.

For notational purposes, we only use the above set of composite
operators when referring to block A. When considering another
block B, we systematically replace $a_i$ by $b_i$.

The key point is that all matrix elements needed for the iterative
algorithm can be obtained from these composite operators. To see
this, consider for example the Hubbard Hamiltonian in a product
space A$\times$B. It can be written as

\begin{eqnarray}
H_{\rm{A}\times\rm{B}}&=&H_A +H_B +\frac{U}{N} \sum_p \left\{
\sum_{\sigma'} [ \frac{1}{4} a_1 (p,\sigma') b_1(-p,\bar{\sigma}')
+   b^{\dagger}_0(p,\sigma') a_3(p,\sigma') \right.
\nonumber \\
&~& \left. + (a \leftrightarrow b) ] ~+~ a^{\dagger}_4(p)b_4(p) -
b_2(p)a^{\dagger}_2(p) \right\}~+~h.c. \label{H_coupled}
\end{eqnarray}
Note that in the A space, we only need the set of $a$ operators
and correspondingly in the B space we only need the analogous set
of $b$ operators.

In more complex spaces, with larger numbers of coupled blocks, the
same remarks can be straightforwardly shown to apply.

The above development is specific to the Hubbard Hamiltonian of
Eq. (\ref{Hub}). However, as we will discuss later, similar
considerations are essential to preserve the computational
feasibility of the DMRG algorithm in the more general quantum
problems that arise in the description of finite Fermi systems.

\subsubsection{Translational invariance and momentum conservation.}

As noted earlier, the real-space DMRG when applied to uniform
quantum lattices benefits greatly from the fact that all positions
on the lattice look the same. Thus, once a certain portion of the
lattice has been treated, the same results can be transferred to
any equivalent set of sites.

The momentum-space DMRG loses this simplicity. A given group of
quantized momentum levels has no obvious relation to any other
group.

Of course the equivalence of a given group of sites on a uniform
lattice to any other analogous group was a consequence of the
translational invariance of the problem. That symmetry is not lost
by transforming to momentum space; it simply shows up elsewhere.
In particular, in momentum space the translational invariance of
the lattice shows up in terms of momentum conservation. When
building the superblock, we need not mix states of different total
momentum. This has the advantage of reducing the size of the
superblocks in momentum space compared to real space, even for the
same assumed block dimension.

In building states of a given total momentum in the superblock, we
must maintain all allowed momentum states for the blocks
themselves. The resulting density matrix for the block will thus
contain different momenta. However, it will be block diagonal in
momentum and thus likewise be effectively reduced in dimension
relative to the corresponding matrix in real space.

Similar considerations play a role in applications of the DMRG to
finite Fermi systems, where symmetries likewise must be treated
carefully. In rotationally-invariant systems, for example, angular
momentum is a conserved quantum number. As we will discuss in
Section IV, full incorporation of angular momentum symmetry can be
built into the formalism, but at an increase in complexity.
Minimally, however, when working in a basis of single-particle
states with well-defined angular momentum projection, we can build
states with conserved m-projection and thereby reduce the size of
the superblocks that must be diagonalized. Any additive quantum
number associated with an Abelian symmetry can be used in this way
to reduce the size of the superblock and to make the reduced
density matrices block diagonal.

\subsubsection{The warm-up phase of the kDMRG algorithm.}

In a real-space description of uniform lattices, we can use the
equivalence of contiguous sites throughout the lattice to
facilitate the introduction of a medium with which to build a DMRG
superblock in the warm-up phase. In momentum-space, as noted
above, this is no longer the case. On the one hand, this makes it
clear that it is not appropriate to use the infinite DMRG
algorithm when working in momentum space. But even with the finite
algorithm it is necessary to adopt a strategy for implementing the
warm-up phase, where the first approximation to the description of
the the various size blocks is obtained.

In the work of Xiang and the others who have implemented the
kDMRG, the procedure has been to perform truncation in the warm-up
phase according to Wilson's numerical RG procedure, namely to
truncate to the lowest eigenstates of the enlarged block
Hamiltonian. As for a real-space description of non-uniform
lattices, the expectation is that subsequent sweeps should correct
any deficiencies that might exist in this procedure.

\subsubsection{Applications of the kDMRG.}

Having described the key elements of the kDMRG method, we now
discuss some of the applications that have been reported.

We first discuss some of the results reported by Xiang in the
paper that first introduced the method \cite{xiang}. The
calculations were carried out for the 1D and 2D Hubbard models,
using the standard Hamiltonian of Eq. (\ref{Hub}). They were
performed as a function of the number of lattice sites $N$ and the
ratio $U/t$ of the strength of the on-site Coulomb repulsion to
that of the one-body hopping term. In the case of the 1D model,
all calculations were done at half filling, whereas the 2D
calculations considered different filling fractions. In all cases,
the number of states $m$ kept in each truncation was 1000 and
comparison with exact results was reported where available.

Applications in 1D showed two important points: (1) that the
momentum-space DMRG method is significantly more accurate than the
Wilson RG when applied to the 1D Hubbard model, and (2) that the
kDMRG results for 1D lattices are less precise than those for
real-space DMRG.

The latter conclusion is worthy of further comment.  As noted
earlier, the accuracy of the DMRG algorithm depends critically on
how rapidly the correlations fall off. In real space, the
correlations decay with distance since the Hamiltonian only
couples neighboring lattice sites. An analogous statement does not
apply in momentum space, where the interaction is highly non-local
and can strongly couple momentum states that are far apart. A
critical question there is how strong is this coupling relative to
the single-particle splittings between the Bloch states.
Nevertheless, the conclusion to draw from this work is that when
studying 1D lattices using the DMRG algorithm, the real-space
formalism remains the method of choice.

Xiang's results for the Hubbard model on a $4 \times 4$ 2D lattice
are summarized in table \ref{Xiang2D}. The results are shown for
several values of $U/t$ and for several numbers of electrons.
Where available they are also compared with results of Quantum
Monte Carlo (QMC) and Stochastic Diagonalization (SQ)
calculations. There are two key points to note. The first is that
the method works very well for weak on-site repulsion. This should
not be surprising, since it is exact when $U=0$. In that case, the
exact solution - a Slater determinant of Bloch states up to the
Fermi momentum - is a trivial one-state kDMRG wave function. The
second point to note is that the method continues to work
reasonably well even for stronger coupling, producing results that
are competitive with those obtained using the two other popular
many-body methods against which it was compared.

\begin{table}
\caption{\label{Xiang2D}Comparison of the ground state energy
calculated for the $4 \times 4$ Hubbard model. The exact results
are compared with those obtained using the kDMRG method, the
Quantum Monte Carlo (QMC) method and the Stochastic
Diagonalization (SQ) method for different numbers of electrons $N$
and different values of $U/t$. From Table II of ref.
\protect\cite{xiang}. Reprinted by permission of the American
Physical Society.  }
\begin{indented}
\item[]\begin{tabular}{cccccc}
\br
$U/t$ & $N$ &  $Exact$  & $kDMRG$ & $QMC$ & $SQ$ \\
\mr
2& 16 & -18.01757 &   -18.012 &   &  \\
4 & 14 & -15.74459 & -15.673 & & \\
4 & 16 & -13.62185 & -13.571 & -13.6 & -13.59\\
8 & 16 & -8.263 & -8.48 & \\
\br
\end{tabular}
\end{indented}
\end{table}

Several years later, Nishimoto and collaborators \cite{NJGN}
reinvestigated the usefulness of the kDMRG method for treating the
Hubbard model. They redid some of the calculations reported by
Xiang and also carried out a much more systematic investigation,
considering several different hopping scenarios. An important
difference of their calculations relative to those of Xiang was in
the blocking scheme. These authors applied White's blocking
scheme, adding two sites at a time to enlarge the superblock.
Xiang, in his kDMRG work, added only one site at a time. This
makes it somewhat difficult to draw comparative conclusions, as
the sizes of the superblocks that would arise in the two blocking
approaches are different for the same number of states $m$
retained in each block.

We summarize here some of the key conclusions of ref. \cite{NJGN}.

\begin{itemize}
\item
In 1D with nearest-neighbor hopping, Xiang's results do not match
theirs. Several possible explanations were put forth, including
that Xiang's calculations may not have converged and that it could
be a consequence of the different blocking schemes used in the two
calculations.
\item
In 2D, both the rDMRG and the kDMRG become less accurate with
increasing dimensionality, albeit for different reasons. In real
space, it relates to the level of off-diagonal coupling. In
momentum space, it relates to the increased degeneracies and to
the long-range nature of the interaction.
\item The kDMRG is less
dependent on the form of the hopping term. In real space, the
method becomes rapidly worse as the hopping range increases; this
is not the case in momentum space.
\item
In agreement with Xiang, they find that the accuracy of the DMRG
method becomes worse with increasing strength $U$, scaled by the
band width.
\item In the case of the $4 \times 4$ 2D lattice at half filling,
the kDMRG was found to be
more accurate than the rDMRG for $U<8t$.
\end{itemize}

\section{DMRG for finite Fermi systems}

The real-space DMRG has proven extremely successful in describing
the physics of one-dimensional systems with short-range
interactions. As discussed in Sect. II.B.4, the method can be
extended to 2D systems by selecting a particular path covering the
lattice at the expense of introducing long-range interactions.
These long-range interactions arise between neighboring sites in
the 2D lattice, which may be located far away within the selected
1D path. The interplay between short-range and long-range
interactions works to the detriment of the optimal accuracy of the
1D DMRG and to an increase in the memory requirements to implement
the algorithm, because of the necessity of storing the matrix
elements of long-range correlators. In practice, real-space 2D
DMRG calculations were therefore restricted to small clusters or
to ladders up to six legs \cite{new}.

The kDMRG represented a possible way to extend the DMRG
methodology to larger dimensions. Unfortunately, the results
obtained for the Hubbard models of 2D lattices were not very
encouraging and the procedure was not pursued very much. However,
the formal procedure developed by Xiang to reduce the number of
matrix elements needed to store at each iteration was a critical
step in the subsequent application of the DMRG methodology to
finite Fermi systems, to which we now turn.

In finite Fermi systems one typically starts with a one- plus
two-body Hamiltonian expressed in some single-particle
basis,\bigskip
\begin{equation}
H=\sum_{ij}T_{ij}c_{i}^{\dagger
}c_{j}+\sum_{ijkl}V_{ijkl}c_{i}^{\dagger }c_{j}^{\dagger
}c_{l}c_{k} ~, \label{H1}
\end{equation}
where $T_{ij}$ takes into account the kinetic energy and the
mean-field interaction between particles and $V_{ijkl}$ represents
the residual interaction.

A typical basis used in studies in quantum chemistry, nuclear
physics, matter waves, etc. involves the use of the Hartree-Fock
(HF) mean-field approximation. Other frequently-considered
possibilities include the Kohn-Sham basis or approximate mean
fields. As an example of the latter, nuclear structure
calculations often use for simplicity a 3D harmonic oscillator
basis or a basis comprised of the eigenstates of a Woods-Saxon
potential, as approximations to the nuclear mean-field potential.

The first attempt to apply the DMRG to a 3D finite Fermi system
was performed by White and Martin\cite{W5}. They proposed the use
of the DMRG for $ab$ $initio$ calculations of the electronic
structure of molecules. The method was further studied and refined
for calculations in quantum chemistry and for application to
quantum Hall systems \cite{shi1,shi2,shi3,shi4,shi5}. Other forms
of the DMRG procedure were developed and applied to ultrasmall
superconducting grains\cite{duke1,duke2} and the nuclear shell
model\cite {duke3,duke4,duke5}. In the following subsections, we
describe the progress achieved in each of these domains, as well
as the difficulties and limitations of the procedures that were
used.

\bigskip

\subsection{Quantum Chemistry}
\label{QC}

\bigskip {\em Ab initio} calculations in Quantum Chemistry (QC) are based
on the approximate treatment of a system of $N$ electrons moving
in a field created by $M$ fixed atomic nuclei and interacting
through the Coulomb force. The electronic Hamiltonian of the
system is

\bigskip
\begin{equation}
H=\sum_{i=1}^{N}-\frac{\bigtriangledown
_{i}^{2}}{2}-\sum_{i=1,~\alpha
=1}^{N,M}\frac{Z_{\alpha }~e^{2}}{\left| r_{i}-r_{\alpha }\right| }+\frac{1}{%
2}\sum_{i\neq j=1}^{N}\frac{e^{2}}{\left| r_{i}-r_{j}\right| } ~.
\label{he}
\end{equation}

There are two possible avenues \cite{QuantumChemistry} for the
treatment of the electronic Hamiltonian (\ref{he}). One involves
the use of density functional theory (DFT), which has been used
extensively and with great success in the treatment of fairly
large molecules. The other involves the use of Hartree-Fock (HF)
mean-field theory as a starting point for an improved treatment of
the electron-electron interaction.

In principle, the best means of treating the electron-electron
interaction in a HF framework is through the full Configuration
Interaction (CI) approach \cite{QuantumChemistry}. In this
approach, the electron-electron interaction is exactly
diagonalized  in a basis of orthogonal states generated by the HF
method. For problems that are not too large, this can be
implemented through the use of diagonalization routines
appropriate to sparse matrices, such as Lanczos or Davidson. With
present computational resources, matrices of dimension $10^{7}$ to
$10^{8}$ can be treated in this way, with little hope of a
dramatic increase in the future. Other approximate but very
accurate methods to treat the residual electronic interactions,
such as Coupled Cluster Theory (CCT), are often used, but like the
full CI they scale exponentially with the number of basis states,
thereby imposing limitations on the Hamiltonian matrix dimensions
that can be treated and permitting complete calculations for
relatively small molecules only.

The DMRG method was proposed by White and Martin \cite{W5,W6} as
an alternative procedure for approximately diagonalizing the
residual electronic interaction, but in a much more
computationally efficient way. Despite requiring a much slower
polynomial computational effort, the method nevertheless has been
shown to provide highly accurate results. More precisely, the DMRG
computational effort scales as $O(n^{4}m^{2})$ where $n$ is the
total number of basis states and $m$ is the number of states
retained in the iterative algorithm. This suggests that
application of the DMRG algorithm to the field of QC could open up
the possibility of a wide range of highly accurate molecular
structure calculations, well beyond the limits of the full CI.

The usual way to generate a second quantized Hamiltonian in a
finite molecular basis is to first consider a small number of
atomic orbital states for each atomic nucleus, generating a local
non-orthogonal basis, and then to use HF mean-field theory to
produce an orthogonal molecular basis of $n$ orbitals,
doubly-degenerate in the spin quantum number. The second-quantized
Hamiltonian in this basis is

\begin{equation}
H=\sum_{ij\sigma }T_{ij}c_{i\sigma }^{\dagger }c_{j\sigma }+\frac{1}{2}%
\sum_{ijkl\sigma \sigma ^{\prime }}V_{ijkl}c_{i\sigma }^{\dagger
}c_{j\sigma^{\prime} }^{\dagger }c_{l\sigma^{\prime} }c_{k\sigma }
~, \label{hes}
\end{equation}

\noindent where $c_{i\sigma }^{\dagger }\left( c_{i\sigma }\right)
$ is the creation (annihilation) operator of an electron in
orbital $i$ with spin projection $\sigma $. The matrix $T$ is a
radial integral of the electron kinetic energy and the
electron-nucleus interaction in the HF basis, while the matrix V
is a corresponding radial integral of the repulsive Coulomb
interaction between electrons.

The key point in applying the DMRG procedure to QC problems is to
treat each orbital as a ``site" in a one-dimensional lattice. This
site can be empty, singly occupied with spin up, singly occupied
with spin down, or doubly occupied. The situation is analogous to
the kDMRG and in fact it was the use of the Xiang trick for
reducing the three-index operators to an effective one-index
operator that made the procedure tractable in QC. The most complex
operators that need to be calculated and stored in each iteration
are those with two electronic creation and/or annihilation
operators, requiring storage space of order $n^{2}m^{2}$.

As we will see below, a key issue, which is still not in our view
fully resolved, is how to establish the optimal relative order
among the orbitals. Two possibilities are to order the orbitals in
terms of HF energies or to maximize the interactions between
neighboring orbitals. A satisfactory resolution of this problem
will ultimately lead to an optimized DMRG protocol for
applications in QC.

\begin{figure}
\begin{center}
\hspace{2cm}\epsfysize=8cm \epsfxsize=9cm\epsfbox{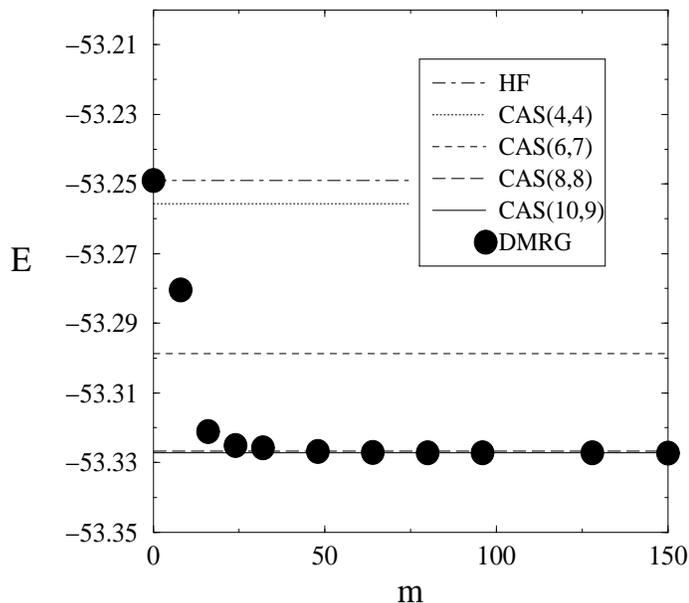}
\end{center}

\caption{Ground-state energy $E$ of CH$_{4}$ as a function of the
number of states retained $m$ at each DMRG iteration. The lines
are the Hartree-Fock and CAS results. Taken from \protect
\cite{W6}. Reprinted by permission of John Wiley \& Sons, Inc.}
\label{ch4}
\end{figure}

As an illustration of the first tests of the method, we show in
fig. \ref{ch4} DMRG results for the methane molecule ($CH_{4}$)
\cite{W6}. The system consists of $10$ electrons moving in a
minimal basis of $9$ orbitals. The one- and two-body matrices in
the Hamiltonian (\ref{hes}) were obtained following a HF
calculation. The HF molecular orbits played the role of sites in a
finite one-dimensional lattice and the finite DMRG algorithm was
applied to the problem, with the ground state (GS) targeted. The
DMRG results for the GS energy are plotted as a function of the
number of states $m$ retained in the iterative procedure. For
comparison, the figure also includes the CAS($\mu ,\nu $) results
corresponding to diagonalization of the Hamiltonian (\ref{hes}) in
a Complete Active Space of $\nu $ valence orbitals and $\mu $
active electrons.

The DMRG results exponentially converge to the exact results of
CAS($10,9$) in which the complete active space of $9$ molecular
orbits is taken into account. For rather small $m$ values
($\sim50$) the DMRG results are already comparable to CAS($8,8$)
in which two electrons are frozen in the lowest HF orbital.

The influence of the choice of basis on the global procedure was
also investigated in \cite{W6}, with a focus on the singlet state
of the H-He molecule. The standard definition of the molecular
orbits based on HF mean field approximation was compared with the
DFT definition of natural orbitals in two different approximations
-- the Local Density Approximation and the Generalized Gradient
Approximation. The results obtained were fairly insensitive to the
choice of basis, showing similar convergence properties.

The efficiency of the DMRG procedure in QC has been further
studied in ref. \cite{q1}. Though the method was found to perform
very well for the molecules Be$_2$ and HF, comparison with full CI
results for the N$_2$ molecule did not produce satisfactory
results. The reasons for this failure were not conclusive, but
there were indications that a non-optimal ordering of orbitals may
have played an important role.

\begin{table}
\caption{\label{head}DMRG results for the water molecule for
various values of  $m$, compared with three levels
 of approximation in CCT. Taken from \protect \cite{q2}. Reprinted by
 permission of the American Institute of Physics.}
\begin{indented}
\item[]\begin{tabular}{lrrr}
\br
 $m$ & $E/H$ & $\delta E/mH$ \\
\mr
 $100$ & $-76.2545$ & $2.1$ \\
 $200$ & $-76.2559$ & $0.71$ \\
 $300$ & $-76.25632$ & $0.32$ \\
 $400$ & $-76.256477$ & $0.157$ \\
 $500$ & $-76.256540$ & $0.094$ \\
 $600$ & $-76.256592$ & $0.042$ \\
 $750$ & $-76.256617$ & $0.017$ \\
 $900$ & $-76.256624$ & $0.004$ \\
 \mr
$CCT$ & $-76.252503$ & $4.131$ \\
$CCT(T)$ & $-76.255907$ & $0.727$ \\
$CCT(TQ)$ & $-76.256846$ & $0.202$ \\
\br
\end{tabular}
\end{indented}
\end{table}

A very careful study of all of the ingredients that define the
performance and efficiency of the DMRG algorithm in QC was
reported in ref. \cite{q2}, where the reader can also find a
detailed and pedagogical presentation of the method. The subject
of the orbital ordering was addressed in this work, with the idea
of minimizing the range of the interaction matrix elements in
order to render the correlation length of the system a minimum.
After trying various prescriptions, the authors came to the
conclusion that the simplest method was to minimize the band-with
of the one-body matrix elements $T_{ij}$. They found that the
symmetric reverse Cuthill-McKee reordering method could bring the
$T$ matrix to a block diagonal form, avoiding long-range hopping
terms.

The example of the water molecule, which had been treated by White
and Martin in \cite{W5} as a first test of the DMRG method in QC,
was re-investigated in \cite{q2}, to compare the accuracy of the
DMRG not only with a full CI calculation but also with current CCT
approximations. This problem involves $8$ electrons moving in $25$
active orbits (the $1s$ orbit was frozen). We reproduce in table
\ref{head} their results. The exact GS energy given by a complete
CI diagonalization is $-76.256634 H$. CCT(T) and CCT(TQ) refer to
Coupled Cluster calculations with triples and with triples and
quadruples, respectively. It can be seen from the results that for
$m=400$ the DMRG already has comparable accuracy to CCT(TQ).
Furthermore, and most importantly, since DMRG is a variational
theory we can be sure that the numerical result it yields is an
upper bound to the GS energy. This is not the case for CCT, as
seen from the CCT(TQ) result for which there is overbinding of
$0.2 \mu H$.

Analysis of larger molecules, such  as $C_2 H_4$, indicates that
it is difficult for the DMRG method to capture long-range electron
correlations because of the sequential procedure it uses for
adding orbitals. Nevertheless, the fact that the DMRG results
obtained in ref. \cite{q2} were comparable to the CCT results
including quadruples suggests that it should be taken seriously in
such calculations, with possible improvements now to be discussed.

\begin{figure}
\begin{center}
\hspace{2cm}\epsfxsize=11.25cm \epsfysize=9cm \epsfbox{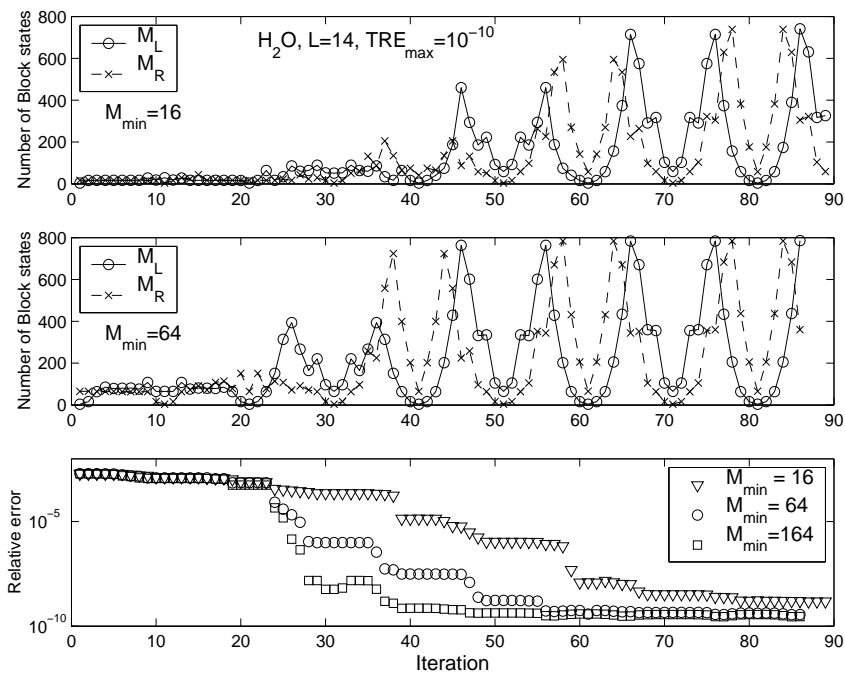}
\end{center}

\caption{\label{lege1}Dynamically selected number of left and
right blocks states for two values of the minimum threshold
$m_{min} =16, 64$ and the relative error as a function of
iteration for three values of $m_{min}= 16, 64, 164$. In all cases
the 10 electrons of the $H_2 O$ molecule were correlated in the
double-zeta water model with 14 orbitals, and $TRE_{max}=10^{-10}$
was set in advance of the calculations. Taken from ref. \protect
\cite{Leg1}. Reprinted by permission of the American Physical
Society.}
\end{figure}

A difficulty in drawing firm conclusions from the studies just
discussed is that exact results from full CI calculations were not
available in all cases, making it impossible to confidently assess
the true accuracy of the DMRG and CCT calculations. One of the
control parameters of the DMRG method is the number of states $m$
retained in the algorithm. The variational properties of the DMRG
assure that the GS energy decreases, or at worst stays constant,
with increasing $m$ values. Indeed, in many cases an exponential
or modified exponential \cite{q2} convergence with $m$ has been
found. Nevertheless, even in such cases it is difficult to
correlate the number of states $m$ with the final precision of the
procedure. In ref. \cite{Leg1}, a new method for determining the
accuracy of the DMRG in advance, the Dynamical Block State
Selection (DBSS) method, was proposed. The truncation error
($TRE$) in each iteration is $TRE=1-\sum_\alpha \omega_\alpha $,
where $\omega_\alpha $ are the eigenvalues of the reduced density
of the block in descending order. Keeping the same number $m$ of
states in each iteration induces fluctuations in the $TRE$, making
it impossible to infer the final accuracy of the procedure. The
DBSS method inverts the procedure for selecting states by defining
a maximum truncation error $TRE_{max}$ at the start, which amounts
to a {\em dynamical} selection of the number of states kept in
each iteration. The DBSS is supplemented by the selection of a
lower threshold $m_{min}$ for the number of states retained. The
first and second panels of figure \ref{lege1}, extracted from ref.
\cite{Leg1}, show the number of states dynamically selected in
each iteration for two values of $m_{min}$ in a calculation for a
water molecule with $10$ electrons and $14$ orbitals. The third
panel shows the relative error in the GS energy, defined as
$(E_{DMRG} - E_{CI})/E_{CI}$.

It is interesting to note that even though the maximum number of
states does not depend on $m_{min}$ the rate of convergence is
faster for larger $m_{min}$ values. In the example of the water
molecule, the HF single particle levels were ordered with
increasing values of energy. In other test cases, {\em e.g.} for
the $CH_2$ molecule \cite{Leg1}, this ordering led to a local
minimum and only by changing the ordering was it possible to
obtain global convergence.

The problem of the single-particle ordering was investigated
recently in ref. \cite{Leg3}, where a protocol for the
initialization of the procedure and for an optimal ordering of the
levels was proposed. Using concepts borrowed from Quantum
Information Theory, it was shown that the optimal ordering
corresponds to locating the levels with maximum entropy at the
center of the chain. This ordering is not always equivalent to the
Cuthill-McKee ordering that had been proposed in \cite{q2}.

There have been several other applications of the DMRG algorithm
in Quantum Chemistry to large polymers
\cite{p1,p2,p3,p4,p5,p6,p7,p8,p9}. These applications have
typically utilized the original real-space DMRG algorithm building
on the fact that polymers can be represented either by a chain or
by another cyclic structure. Since our focus is on applications
that require the machinery appropriate to finite Fermi systems, we
will not discuss these applications here.

\subsection{Ultrasmall superconducting grains}

In 1959, Anderson \cite{An} posed the question of what would be
the lower size limit for a metallic grain to sustain
superconductivity. Since the average level spacing $d$ in a
metallic particle is inversely proportional to its volume, for
small enough particles (namely when $d$ is of the order of the
bulk gap $\Delta$) the scattering of pairs across the Fermi
surface ceases to be an efficient mechanism for lowering the
energy of the system. At this point superconductivity should
therefore disappear. Anderson estimated the critical grain size to
be in the range of nanometers. This old question has recently been
revived in the context of experiments performed by Ralph, Black
and Tinkham \cite{RBT} on ultrasmall Aluminum grains. They
succeeded in constructing a Single Electron Transistor whose
central island was an Al grain of a few nanometers with an
estimated mean level spacing $d\sim 0.4 mev$, of the order of the
bulk gap. The experiments showed the existence of a spectroscopic
gap that could be driven to zero by the application of magnetic
fields.

The Hamiltonian proposed in \cite{delft1} to study the physics of
such small grains is of the reduced BCS form,

\begin{equation}
H = \sum_{j=1}^{\Omega} \sum_{\sigma=\pm}(\epsilon_j -\mu)
c^\dagger_{j\sigma} \; c_{j\sigma} - \lambda d
\sum_{i,j=1}^{\Omega} \; c^\dagger_{i+} c^\dagger_{i-} c_{j-}
c_{j+} ~, \label{D1}
\end{equation}

\noindent where $i,j=1,2,\dots ,\Omega $ label the single-particle
energy levels, which are assumed to be equally spaced ($\epsilon
_{j}=jd$), and $c_{j\sigma }$ are the electron annihilation
operators associated with the two time-reversed states
$|j,\sigma=+\rangle$ and $|j,\sigma =-\rangle$. Finally, $\mu $ is
the chemical potential and $\lambda $ is an adimensional BCS
coupling constant, whose appropriate value for Al grains is 0.224
\cite{BvD}. The systems described by this Hamiltonian have been
treated at half filling for which the number of pairs $N$ equals
the number of levels $\Omega$.

It soon became clear that the standard BCS approximation, which so
accurately describes superconductivity for sufficiently large
systems, is not well suited to treat the systems of physical
interest, for which $N\sim 100$. In such systems, there are strong
number fluctuations proportional to $\sqrt{N}$. Several levels of
approximations were thus applied to the Hamiltonian (\ref{D1}), to
see whether any could describe the physics of pairing in finite
systems. In increasing order of complexity, they included i) BCS
approximation with parity projection \cite{delft1}, ii) Lanczos
diagonalization up to $\Omega=23$ with finite-size corrections
\cite{Mas}, and iii) number-projected BCS \cite{delft2}
approximation.  All of these approximations predicted a phase
transition from superconducting behavior to a normal state at a
critical grain size, and in all cases the transition was {\it
abrupt}. In contrast, the DMRG method -- when applied in a
somewhat modified form \cite{duke1} -- showed that the crossover
as a function of decreasing grain size had a smooth reduction in
pairing correlations.

Following completion of \cite{duke1}, the authors learned that the
Hamiltonian (\ref{D1}) had been solved exactly by Richardson
during the 1960s \cite{Rich}. This made it possible to carry out a
comparison of the DMRG results with those obtained from
Richardson's exact solution. For the largest system treated, with
400 doubly-degenerate levels, the condensation energy was
$E^C_{DMRG}=-22.5168$ with a predicted relative error of
$10^{-4}$. Richardson's exact result was
$E^C_{exact}=-22.5183141$, confirming the accuracy of the phDMRG.
and showing that the relative error predicted by the DMRG method
was in fact appropriate. On this basis, it is clear that the DMRG
method provides an essentially exact description of the physics of
pairing in finite systems and that the correct crossover from
superconducting to the normal state is indeed smooth and not
abrupt.

The modifications implemented in the DMRG method for its
application to small metallic grains were based on the fact that
in a finite Fermi system the Fermi energy ($E_F$) naturally
separates the single-particle space into a primarily occupied
subspace and another primarily empty subspace. The method that
emerges from such a construction was called the particle-hole DMRG
(phDMRG), which we now briefly describe.

\begin{figure}
\begin{center}
\hspace{1.5cm}\epsfxsize=5.5cm
\epsfysize=7.2cm\epsfbox{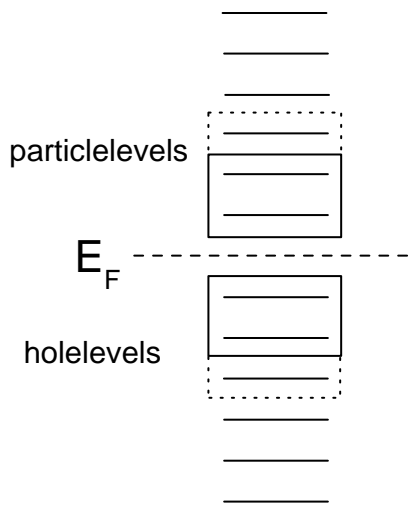}
\end{center}
\caption{\label{fig2}Schematic illustration of the phDMRG method
for finite Fermi systems.  The single-particle levels are divided
into primarily occupied hole levels and primarily empty particle
levels. The first two levels above the Fermi energy $E_F$ have
been formed into a particle block and the first two below have
been formed into a hole block. The third levels above and below
are being added to form enlarged blocks.}
\end{figure}

Consider the set of single-particle levels shown in fig.
\ref{fig2}. Note that they are divided into two sets. The particle
levels are those above the Fermi surface, which should be
essentially empty. The hole levels are those below the Fermi
surface and they should be essentially filled. In the phDMRG
method, one first defines particle and hole blocks from the levels
nearest to the Fermi surface, and then gradually grows those
blocks by adding levels progressively further away. The figure is
meant to signify that we have already created particle and hole
blocks involving the first two available levels, respectively, and
are enlarging them by adding the third level(s). The particle
block then serves as the medium for the hole block and the hole
block as the medium for the particle block as we implement the
DMRG truncation strategy. This process is continued until all
particle and hole levels have been treated.

The phDMRG procedure, as just described, is related to the
infinite DMRG algorithm in that sweeping is not implemented. It is
extremely difficult to implement sweeping with particle and hole
blocks while at the same time preserving the four block structure
of the method. As such, it is expected that the phDMRG will only
work well for systems where the correlations fall off rapidly as
we move away from the Fermi surface. The key point is that at the
beginning of the iterative procedure the levels that are outside
the particle-hole superblock do not impact on the selection of the
optimal state(s) with which to truncate the two blocks. This is a
recurring limitation of the infinite algorithm, which can be
partially overcome by combining the DMRG truncation strategy with
the use of effective interaction theory. In the application to
ultrasmall superconducting grains, this was implemented through a
phenomenological renormalization of the pairing strength at each
iteration. When the system described by the pairing Hamiltonian
(\ref{D1}) for $\Omega$ levels at half filling is considered in
BCS approximation, the bulk gap is given by $\Delta_\Omega = d
\Omega /2 \sinh(1/ \lambda)$. The effect of the levels outside a
superblock that consists of $2n$ levels can thus be taken into
account in an average way by introducing an effective coupling
strength $\lambda_n$ -- corresponding to step $n$ of the iteration
-- such that the gap in the superblock being considered equals the
bulk gap. This is guaranteed if the renormalized coupling strength
$\lambda_n$ is chosen according to

\begin{equation}
{\rm sinh}  \lambda_n = \frac{2(n+1)}{\Omega} {\rm sinh} \lambda
~. \label{64}
\end{equation}
Note that the pairing strength assumes its full value when all
levels are included, {\em i.e.,} when $n=\frac{\Omega-1}{2}$.

The effect of renormalizing the coupling strength in this way can
be seen in table \ref{Renorm}, where we show results for the
condensation energy in units of the mean level spacing, without
renormalization $E^C _{bare}$ and with renormalization $E^C
_{dressed}$, for different values of the number of states $m$
maintained. The system considered has $\Omega=100$ at half filling
and the bare coupling constant is $\lambda=0.4$, for which the
system is well into the superconducting phase. In the last column,
we show the dimension of the largest superblock required in the
iterative process, which should be compared with the total
dimension of the problem, $10^{29}$ .

\begin{table}
\caption{\label{Renorm}The effect of renormalization on the
results of phDMRG calculations for ultrasmall superconducting
grains.}
\begin{indented}
\item[]\begin{tabular}{|c|c|c|c|}
\br
$m$ & $E_{{\rm bare}}^{C}/d$ & $E_{{\rm dressed}}^{C}/d$ & ${\rm dim%
}{\cal H}_{100,m}$ \\ \mr
40 & -40.32502 & -40.49884 & 1246 \\
50 & -40.44623 & -40.50014 & 2108 \\
60 & -40.46887 & -40.50061 & 3032 \\
70 & -40.48878 & -40.50068 & 3622 \\
80 & -40.49588 & -40.50072 & 4820 \\
90 & -40.49815 & -40.50074 & 6306 \\
100 & -40.49919 & -40.50075 &  7778 \\
110 & -40.49983 & -40.50075 &  9720 \\
\br
\end{tabular}
\end{indented}
\end{table}

\begin{figure}
\begin{center}
\hspace{2cm}\epsfxsize=10cm \epsfysize=6.7cm \epsfbox{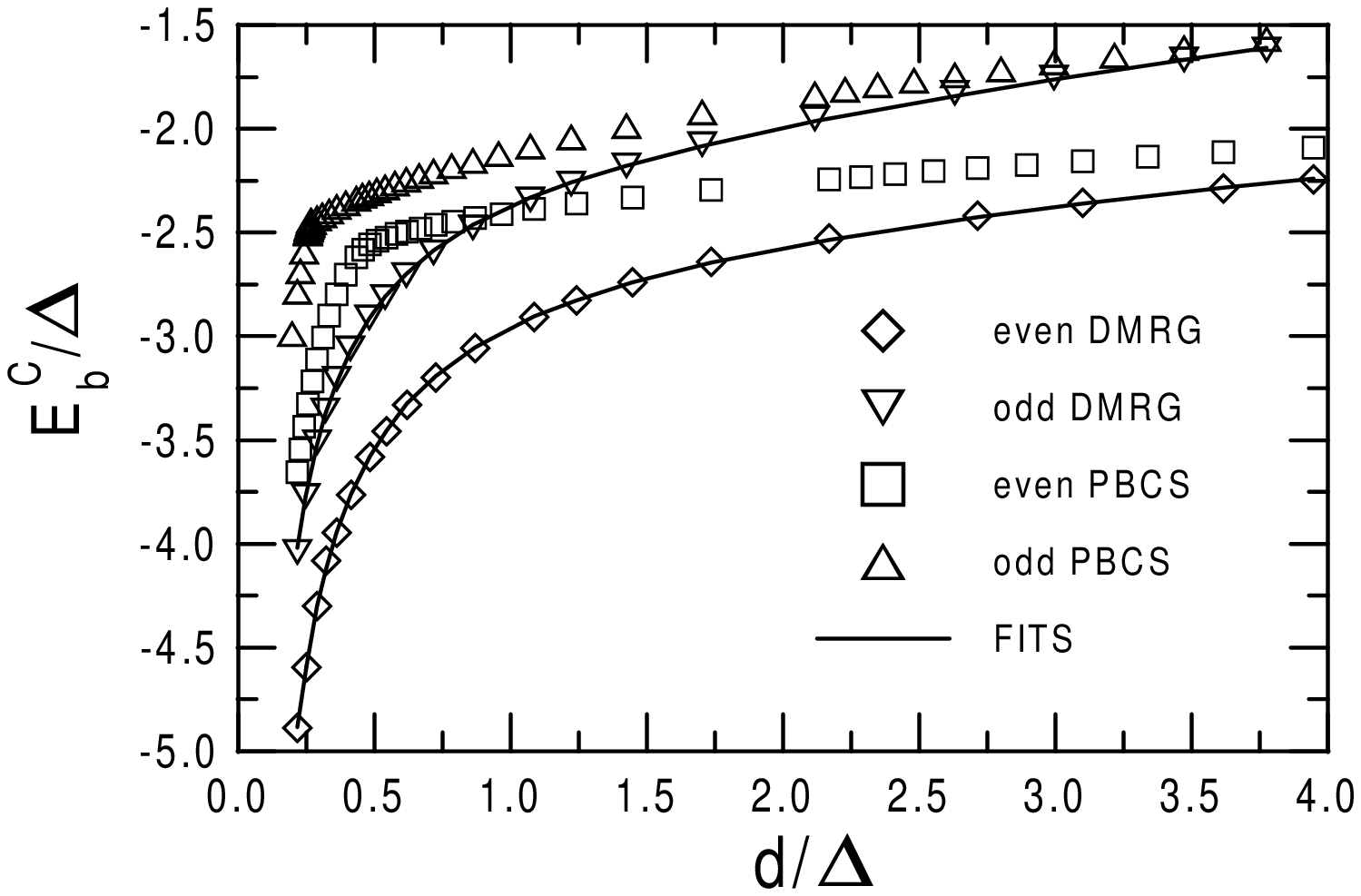}
\end{center}
\caption{\label{fig1}Ground state condensation energies $E^C_b
(b=0,1)$ as a function of $d/\Delta$ for $\lambda=0.224.$.
$\Omega$ ranges from 22 ( 23) up to 400 (401) for even (odd)
grains and $m=60$. The PBCS results are those of ref.
\protect\cite{delft2}. Taken from \protect \cite{duke1}. Reprinted
by permission of the American Physical Society. }
\end{figure}

As can be seen from the table, there is a significant improvement
in convergence when renormalization of the coupling constant is
implemented in the phDMRG algorithm. For $m=100$, for which the
largest matrix diagonalized had a dimension of roughly $8,000$,
the condensation energy (column $3$) has converged to seven
significant figures, while the standard phDMRG procedure (column
2) without renormalization is still far from convergence. The
results of the complete calculation for ultrasmall superconducting
grains are shown in fig. \ref{fig1}. The condensation energies for
even and odd systems are displayed as a function of the mean level
spacing in units of the bulk gap. The solid lines correspond to
the numerically exact phDMRG calculation and the squares and
triangles, for even and odd systems, respectively, correspond to
the PBCS calculations of \cite{BvD}. The phDMRG results were
obtained with $m=60$  with an estimated relative error for the
largest system considered ($\Omega=400$ levels) of $10^{-4}$. The
phDMRG results are consistently below the PBCS results,
representing a considerable gain in correlation energy. Most
importantly, while the PBCS results suggest an abrupt change of
the condensation energy at a critical value of the level spacing,
the phDMRG results unveiled the true nature of the transition.
There is a very smooth crossover as the size of the grains
decrease, from superconducting grains to metallic grains with
strong pairing fluctuations.

The fact that BCS approximation ceases to provide a good
approximation for small enough grains suggests that other
phenomena associated with superconductivity may also be strongly
modified in that regime. One example is the Josephson effect,
whereby correlated Cooper pairs tunnel from one superconducting
system to another due to the differences in phases of their
respective order parameters. When BCS approximation breaks down,
it is no longer meaningful to speak of superconducting phases,
raising the question of what is the fate of the Josephson effect
between two small superconducting grains. This question has
recently been addressed in ref. \cite{delft3}.

To study the physics of two very small grains, it is necessary to
have a computational framework that can accurately describe the
dynamics. From what we discussed earlier, BCS approximation is
inadequate in a regime in which the two grains are sufficiently
small. The Richardson approach, which gives the exact solution for
individual grains, no longer applies when there are two distinct
grains with tunnelling between them.  The phDMRG approach, which
had been used successfully for individual grains, can be naturally
generalized to a two-grain scenario and might therefore provide a
useful framework for an accurate treatment of the tunnelling
phenomenon. In ref. \cite{delft3}, the two-grain phDMRG method was
developed and applied to this problem. The basic idea of the
method is to add a particle level and a hole level for one type of
grain and then to follow that by adding a particle level and a
hole level of the other type, gradually moving away from the Fermi
surface. At each step, truncation is implemented in the particle
block (which includes levels from both grains) and the hole block
(likewise with levels from both grains) based on the associated
reduced density matrices. This process is continued until all
levels have been treated.

There are two variables that define the two-grain problem. One is
the size of the grains, or equivalently the number of
single-particle levels that contribute. The second is the
Josephson coupling, namely the strength of the term in the
Hamiltonian that governs pair tunnelling.

\begin{figure}
\begin{center}
\hspace{2cm}\epsfxsize=9cm \epsfysize=6.6cm\epsfbox{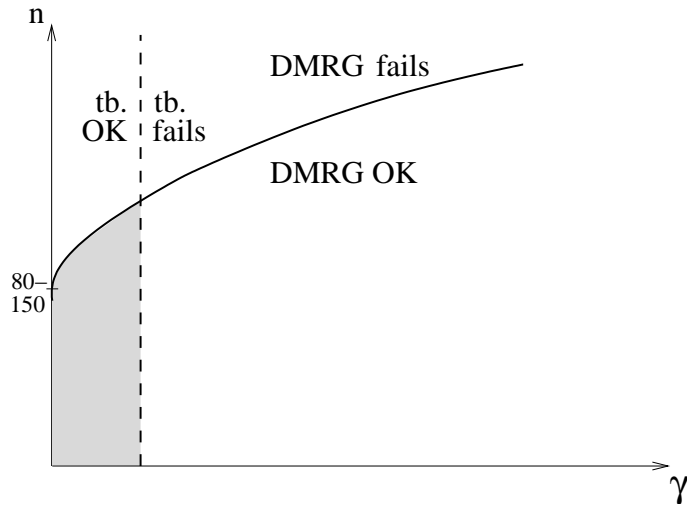}
\end{center}
\caption{\label{2grain}Schematic illustration of the regimes of
validity of the two-grain DMRG method and the tight-binding (tb)
method in the description of tunnelling between two
superconducting grains. $\gamma$ represents the inter-grain
coupling and $n$ the number of energy levels. Taken from Ref.
\protect\cite{delft3}. }
\end{figure}

For the purposes of this review, perhaps the key point that
emerged from the study concerns the range of validity of the
two-grain DMRG method as a function of these two variables. This
is schematically illustrated in fig. \ref{2grain}, together with
an analogous illustration of the range of validity of the
tight-binding (tb) method.

The tight-binding method -- often called the weak-coupling method
-- assumes that the coupled system can be represented as a product
of the ground states of two decoupled systems, which themselves
can be obtained from the ordinary one-grain DMRG method. As the
figure indicates, this should only work well for very
weakly-coupled grains.

The two-grain DMRG method has a very different range of validity.
It does not in general work well for weakly-coupled grains, except
if the grain sizes are sufficiently small. This can be understood
from the following argument. The DMRG method, in all of its
manifestations, depends critically on the presence of correlations
between the block and the medium in order to effectively truncate
the space. When the two grains are very weakly coupled, there is
no natural way to truncate the block, as all states are of
comparable importance. Of course, if the number of levels is
sufficiently small, the number of states $m$ required to
adequately describe it is also very small and thus the DMRG can
still well approximate it. These points are all nicely borne out
in the calculations reported in ref. \cite{delft3}.

\section{2D Electrons in a High Magnetic Field}

Electrons in high magnetic fields are constrained to move in an
effective two-dimensional space characterized by the Landau
orbits. Depending on the filling of the orbits, various phases
have been predicted and subsequently observed. The Fractional
Quantum Hall (FQH) effect, discovered for various fractional
fillings in the lowest Landau level, was followed by the
observation of different exotic phases, like stripes, bubbles,
Wigner crystallization, etc...

An important understanding of these phases has been obtained by
exact diagonalization for small systems \cite{Stone} and by
Quantum Monte Carlo studies \cite{Mitra}. Clearly, there are
strong limitations to both methods. Large scale diagonalization
can only handle systems with $N_e \sim 10$ electrons, while Monte
Carlo calculations only provide the bulk properties of the ground
state, with the accuracy of its results being limited by the
well-known sign problem.

The DMRG offers here a great opportunity to go well beyond these
limits in the study of 2D electrons in the lower Landau levels for
different filling factors. As in other applications of the DMRG to
2D systems, the main problem is the mapping onto an effective 1D
path. In ref. \cite{shi1}, the authors proposed the use of the
eigenstates of free electrons in a perpendicular magnetic field in
the Landau gauge to represent the local orbitals that play the
role of the single-electron states. The Hamiltonian representing
the Coulomb repulsion among the electrons in the local orbital
basis is written as

\begin{equation}
H=S\sum_{n}c_{n}^{\dagger }c_{n}+\frac{1}{2}%
\sum_{n_{1}n_{2}n_{3}n_{4}}A_{n_{1}n_{2}n_{3}n_{3}}c_{n_{1}}^{\dagger
}c_{n_{2}}^{\dagger }c_{n_{4}}c_{n_{3}} ~, \label{HQ}
\end{equation}%
where $S$ is a classical Coulomb energy and $c_{n}^{\dagger }$
creates an electron in the single-particle state $n=1,... ,M$, $M$
being the total number of single-particle states in the unit cell,
and  $A$ is the interaction matrix of the Coulomb force (for
details see \cite{shi4}).

The Hamiltonian (\ref{HQ}) represents a strongly interacting
electron system with degenerate single-particles states. While the
Landau basis provides a natural mapping to a 1D system by choosing
the single-particle states $n$ as the effective sites in a 1D
lattice, the complete degeneracy of the states without the
existence of a Fermi energy impedes the separation into particle
and hole blocks and the identification of a region of the
single-particle space with strong correlations. Furthermore, the
interaction matrix $A$ is ``long range" in the single-particle
space, in the sense that it has finite matrix elements between
distant sites in the 1D lattice.

The finite algorithm DMRG was applied to the system described by
the Coulomb Hamiltonian (\ref{HQ}) in the basis of Landau orbits
$N=0,1~$and $2$, to study the ground state properties and to
determine the phase diagram as a function of the filling factor
$\nu={N_e}/{M}$ \cite{shi3,shi5}.

\begin{figure}
\hspace{4cm}\epsfxsize=9cm \epsfysize=6.5cm\epsfbox{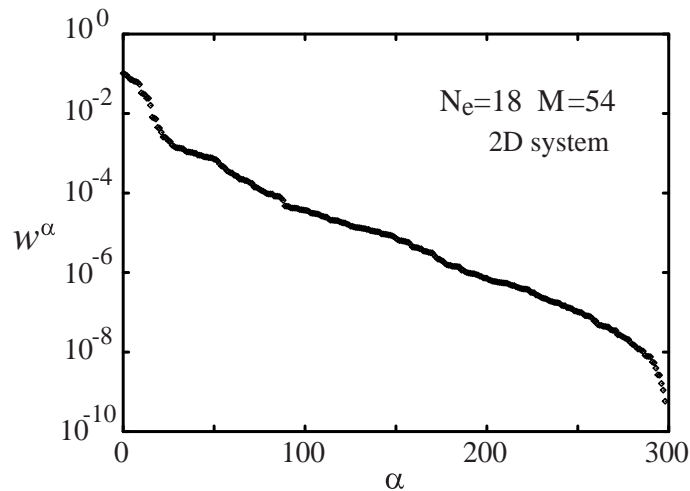}
\caption{\label{shiba1}Eigenvalues $w^\alpha$ of the density
matrix for a system with 18 electrons on 54 local orbitals. Taken
from \protect \cite{shi4}. Reprinted by permission of IOP
Publishing Ltd.}
\end{figure}

In the warm up, the first left block was constructed from the
states $n=1,2$ and the first right block from the states
$n=M,M-1$. The sites with $n=3$ and $n=M-2$ were subsequently
added to the left and right blocks, respectively, in the first
iteration, with a truncation based on the eigenvalues of the
density matrix implemented to retain $m$ states. The method
continued in this way until all sites were taken into account,
completing the warm up. Then, several sweeps were performed until
convergence in the ground-state energy was obtained. A typical
example of the exponential decrease of the density matrix
eigenvalues is given in fig. \ref{shiba1} for a system of $18$
electrons in $M=54$ single-particle states. Retaining $m=200$
states in each iteration leads to a global accuracy of $10^{-4}$.

To identify the properties of the ground state for the different
filling factors,  the expectation value of the pair correlation
function in the DMRG wave function,

\begin{equation}
g\left( \mathbf{r}\right) \equiv \frac{L_{x}L_{y}}{N_{e}\left(
N_{e}-1\right) }\left\langle \Psi \right\vert \sum_{i\neq j}\delta
\left( \mathbf{r}+\mathbf{R}_{i}-\mathbf{R}_{j}\right) \left\vert
\Psi \right\rangle ~,
\end{equation}
was used. Here, $\mathbf{R}_{i}$ is the guiding center of the
$i$th electron, $\mathbf{r}$ is the relative distance between the
pair of electrons and $L_{x,y}$ are the sides of the unit cell.

\begin{figure}
\begin{center}
\hspace{2cm}\epsfxsize=10cm \epsfysize=2.5cm\epsfbox{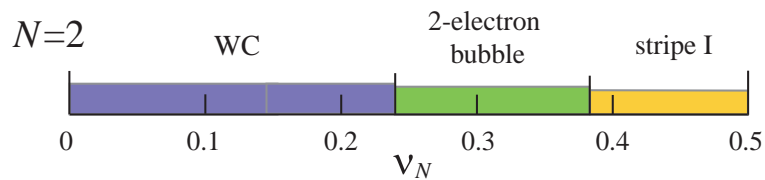}
\end{center}\caption{\label{shiba2}The ground-state phase diagram
of the third-lowest Landau level obtained from a DMRG calculation
with up to $25$ electrons. Taken from \protect \cite{shi5} }
\end{figure}

\begin{figure}
\begin{center}
\hspace{6cm}\epsfbox{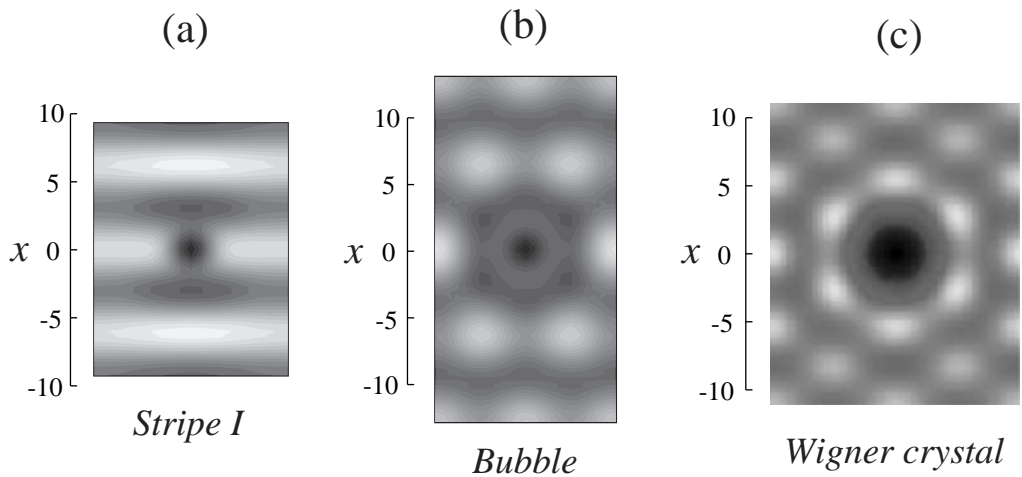}
\end{center}\caption{\label{shiba3}Pair correlation function of
(a) a stripe state, (b) a two-electron bubble state, and (c) a
Wigner crystal. Taken from \protect \cite{shi5}. }
\end{figure}

A thorough study of the phase diagram of 2D electrons in the
lowest three Landau levels was carried out in \cite{shi3,shi5}. We
present in fig. \ref{shiba2}, as an example, the phase diagram of
the third Landau level ($N=2$) \cite{shi5}.  The diagram shows a
Wigner Crystal for low filling factors up to $\nu=0.24$, after
which a transition to a 2-electron bubble phase takes place. There
is a second transition to a stripe phase at $\nu=0.38$. The
properties of the ground state are nicely displayed in fig.
\ref{shiba3}, where we can see the quite marked geometrical
characteristics of the three phases in the pair correlation
function.

The finite algorithm DMRG in the local Landau basis has been shown
to constitute a powerful tool for investigating the physics of 2D
electrons in the lowest Landau levels. Further optimization of the
method could allow the study of larger systems and the
investigation of the behavior of the systems close to the phase
transitions.

\subsection{The DMRG in nuclear structure}

The successes of the DMRG method in Quantum Chemistry, in the
treatment of ultrasmall superconducting grains and in the study of
2D electronic systems suggests that it might also prove useful in
the treatment of another finite Fermi system, the atomic nucleus.
The typical shell-model problem in nuclear structure involves
solving the Schr\"{o}dinger equation with a one- and two-body
interaction in a spherical shell-model basis. The system usually
involves two kinds of interacting particles, neutrons and protons,
with a Hamiltonian

\begin{equation}
H=H_p+H_n+V_{pn}
\end{equation}
that includes three terms -- one that acts solely in the proton
sector, one that acts solely in the neutron sector, and the
proton-neutron interaction.

For heavy enough nuclei and for nuclei with a fairly large number
of active (or valence) nucleons, exact diagonalization of the
shell-model Hamiltonian is not feasible. As an example, to treat
the $J^{\pi}=2^+$ states in the deformed nucleus $^{154}Sm$
assuming an inert $^{132}Sn$ core, 12 valence neutrons in the
orbits
$3f_{7/2},~3p_{3/2},~1h_{9/2},~1i_{13/2},~3p_{1/2},~2f_{5/2}$ and
8 valence neutrons in the orbits
$1g_{7/2},~2d_{5/2}~,1h_{11/2},~2d_{3/2},~3s_{1/2}$ requires
diagonalization in a basis involving $3.46\times10^{14}$ states,
well beyond the limits of the best diagonalization routines. Might
it be possible, however, to iteratively obtain accurate
approximate solutions to the shell-model problem using the DMRG
strategy? This question has been discussed in several recent works
based on the particle-hole algorithm, as we now briefly review.

A characteristic feature of nuclear structure is the presence of
several competing collective structures,  most importantly those
associated with the pairing and quadrupole fields. In refs.
\cite{duke3,duke4}, the phDMRG was applied to a test model
involving both pairing and quadrupole collectivity to assess
whether the algorithm is able to handle their interplay. For
simplicity, the calculations were carried out in a model involving
one type of nucleon in a large single-j shell, interacting via a
$P+Q$ Hamiltonian

\[
 H= - \chi Q \cdot Q - g  P^{\dagger}~ P -
\epsilon \sum_{m} |m| ~ a^\dagger_{jm} a_{jm} ~, \label{H}
\]
where
\[
Q_\mu=\sum_{m} \langle j m+\mu | Q_{\mu} | j m \rangle
c^{\dagger}_{m+\mu} c_{m}
\]
and
\begin{equation}
P^{\dagger}=\sum_ m (-)^{j-m} c^{\dagger}_{m} c^{\dagger}_{-m} ~.
\end{equation}

The third term in the Hamiltonian splits the levels of the
single-j shell into a set of equally-spaced levels of {\em oblate}
character, with the largest $|m|$ value lowest. Because of that
term, the Hamiltonian is not in general rotationally invariant and
thus its eigenstates do not conserve angular momentum. This
unphysical assumption was introduced so that each single-particle
orbit would be doubly degenerate, greatly facilitating the use of
an iterative computational algorithm in the test calculation.

\begin{figure}
\begin{center}
\hspace{2cm}\epsfxsize=6cm\epsfysize=7.5cm\epsfbox{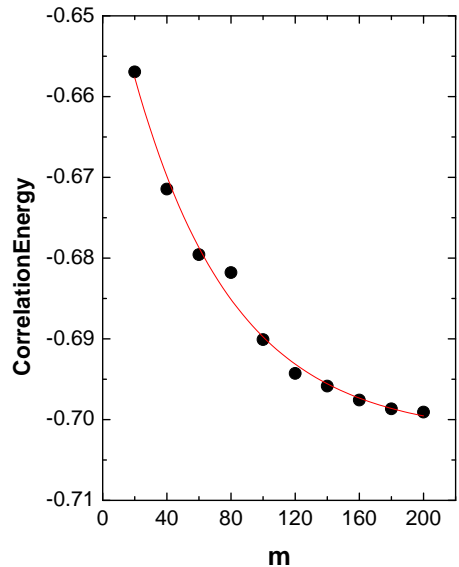}
\end{center}
\caption{\label{j25}DMRG correlation energies as a function of $m$
for a system of 10 particles in a $j=25/2$ orbit subject to a
Hamiltonian with $\chi=1$, $g=0.1$, and $\epsilon=0.1$. The solid
line represents an exponential fit to the DMRG results. Taken from
\protect \cite{duke4}. Reprinted by permission of the American
Physical Society.}
\end{figure}

The first results reported were for a case of $10$ particles in a
$j=25/2$ orbit. The calculations were restricted to states with
total angular momentum projection $M=0$, for which the model
admits 109,583 states, well within the limits of standard
diagonalization routines. Thus, for this problem comparison of the
DMRG results with the exact results was feasible. The results we
show were obtained for the case of $\chi=1,~g=0.1,~\epsilon=0.1$,
for which Hartree Fock Bogolyubov (HFB) approximation shows a
well-defined superconducting solution. As such, the test
Hamiltonian describes a scenario in which pairing and quadrupole
correlations indeed compete.

Figure \ref{j25} compares the correlation energies, defined as the
gain in ground-state energy relative to Hartree Fock
approximation, from the DMRG method and exact Lanczos
diagonalization. For comparison, the exact and HFB values are
$E_{corr}^{Exact}= -0.70633 ~and~ E_{corr}^{HFB} =-0.20641$. By an
$m$ value of $200$, the DMRG method produced more than 99\% of the
exact correlation energy, a dramatic improvement over HFB
approximation which only achieved 28\%. The largest superblock
matrix that had to be diagonalized for this value of $m$ had a
dimension of $2,886$.

Though the calculations used to generate fig. \ref{j25} only
targeted the ground state, they produced very accurate results for
low-lying excited states as well. For $m=200$, the lowest three
excitation energies were reproduced to 1\% or better.

Reference \cite{duke4} also reported results for a much larger
problem, involving 40 particles in a $j=99/2$ orbit. In this case,
the total number of $M=0$ states was $3.84 \times 10^{25}$, much
too large for exact diagonalization. Here too the results were
found to converge exponentially, and the estimated error for the
ground state correlation energy at $m=100$ was just one part in
$10^4$.

\begin{figure}
\begin{center}
\hspace{2cm}\epsfxsize=6cm \epsfysize=6.7cm\epsfbox{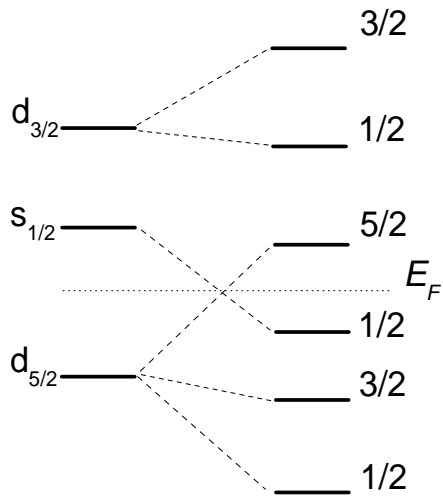}
\end{center}
\caption{\label{fig3}Schematic illustration of the splitting of
the spherical single--particle levels of the $2s-1d$ shell into a
set of doubly--degenerate levels by an axially--deformed
Hartree--Fock calculation. The dashed line represents the Fermi
energy ($E_F$), which separates the particle levels from the hole
levels. Each doubly--degenerate level is labelled by its angular
momentum projection on the intrinsic z-axis. Taken from ref.
\protect \cite{duke5}.}
\end{figure}

The first reported attempt at a realistic application of the
phDMRG method in nuclear structure was presented in ref.
\cite{duke5}. In that work, the nucleus $^{24}Mg$, with four
neutrons and four protons outside doubly-magic $^{16}O$, was
studied. As is usual in the nuclear shell model, the $1s$ and $1p$
core orbits of $^{16}O$ were assumed to be filled and inert and
the remaining 8 nucleons were restricted to the orbits of the
$2s-1d$ shell. This shell-model problem is small and easily solved
by exact shell-model diagonalization.

The results reported in ref. \cite{duke5} were obtained using two
different prescriptions for the ordering in which single-particle
levels were included in the phDMRG algorithm. Here we show the
results obtained using an axially-symmetric Hartree Fock
prescription, with the active levels illustrated in fig.
\ref{fig3}. The axially-symmetric HF single-particle levels are
all doubly degenerate, thereby facilitating the use of the usual
iterative DMRG scheme for adding levels. The calculations were
done using the so-called USD effective Hamiltonian, which has been
tailored to nuclei in the $2s-1d$ shell. Note that the DMRG method
employed in these calculations is closely related to the two-grain
DMRG reported in \cite{delft3} and discussed in Sect. III.B.

\begin{figure}
\begin{center}
\hspace{2cm}\epsfxsize=5.5cm \epsfysize=11cm \epsfbox{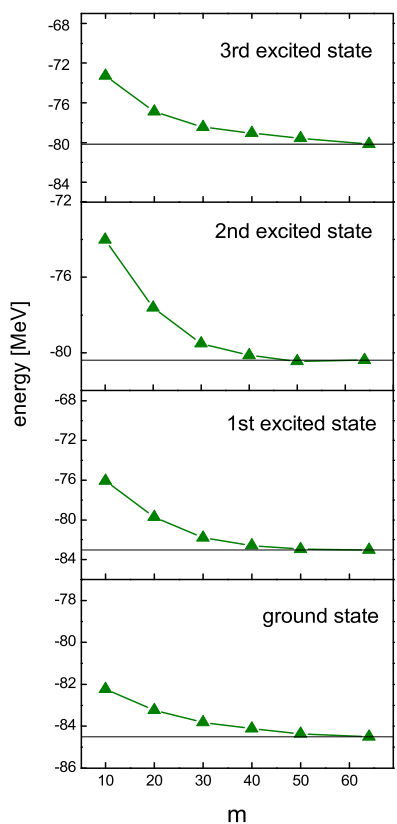}
\end{center}
\caption{\label{24Mg}The calculated energies of the ground state
and the three lowest excited states of $^{24}Mg$ as a function of
$m$, the number of states retained in each block. The horizontal
solid lines refer to the exact energies for the four states. Taken
from ref. \protect \cite{duke5}. }
\end{figure}

Figure \ref{24Mg} shows the results for the energies of the lowest
four states of the nucleus.  For this problem the largest $m$
value that can be realized is $m=64$. The key point to note is
that the convergence to the exact result (shown by the solid line)
is very slow. A value of $m$ above $40$ is required to get
acceptably accurate reproduction of the exact results.
Unfortunately, for such values of $m$, the  matrices that must be
treated are not much smaller than those of the full problem.
Clearly, the phDMRG, method, as implemented in ref. \cite{duke5},
does not work very well for $^{24}Mg$.

There are several possible reasons for the failure of these
calculations. One is that the infinite algorithm that was used
cannot capture the correlations between the different
single-particle levels in the problem. Unfortunately, it is
difficult to incorporate sweeping in the phDMRG, especially for
systems with active neutrons and protons, because of the
preponderance of blocks involved. A second is that the method, as
used, does not preserve rotational invariance. In the following
section, we discuss how rotational invariance, and other
non-Abelian symmetries, can be built into the DMRG. A third
concerns the order in which the levels were included. As noted in
sect. \ref{QC}, a HF ordering, or indeed any ordering based solely
on single-particle energy considerations, may not be optimal.
Finally there are possible issues regarding implementation of the
DMRG truncation algorithm. In particular, no effort was made in
the calculation of ref. \cite{duke5} to ensure that at least one
state of each partition (number of particles/holes and their spin
projection) was maintained in each iteration. In a  more
traditional DMRG study of $^{24}Mg$, it was found that
implementation of such a condition dramatically improved
convergence of the results \cite{PD}.

\section{\protect\bigskip Restoration of Symmetries}

\subsection{General remarks}

Symmetries are a crucial element of all quantum systems. The
invariance of a system under a group of symmetry transformations
gives rise to conserved quantum numbers for the physical states of
the system and associated degeneracies. In the case of uniform
quantum lattices, translational invariance is a natural symmetry,
with the associated conserved quantum number being the linear
momentum. In spin lattices, the total spin of the system is
likewise a conserved quantum number for all physical states. And
in finite Fermi systems, total angular momentum is conserved as a
reflection of the rotational invariance of the system.

Methods of approximately solving the Schr\"{o}dinger equation
frequently break the symmetries of the Hamiltonian and thereby
lead to states that do not contain the appropriate quantum
numbers. Often this is done intentionally, as spontaneous breaking
of symmetries is a natural way to include collective correlations
between the particles in the system. It is then critical, however,
to restore the broken symmetries in order to establish contact
with the states of the physical system, i.e., those probed in
experiment.

The DMRG method, being a truncation of the full Hilbert space of
the problem, in general violates symmetries, {\em unless it is
implemented very carefully}. To understand this, consider the
problem of a spin lattice, where we gradually add spin sites as we
grow the system. To construct states of good total spin, it is
necessary to include all the spin projections required by the
Clebsch Gordan series. Truncation to a limited number of states
based solely on density matrix considerations will not in general
keep all the spin configurations needed to reconstruct the full
spin eigenstates.

As pointed out by White \cite{W4}, it is possible to preserve spin
symmetry, in principle, by careful implementation of the density
matrix strategy. In particular, as long as the density matrix
truncation always includes all states from degenerate multiplets,
it will preserve spin symmetry. However, because of the numerical
limitations inherent in computational algorithms, it is extremely
difficult to guarantee that we do not eliminate states that in a
perfect calculation were degenerate with those that are kept.
Equally important, it is not feasible to fix the number of states
kept a priori in such a calculation, since that number is dictated
by the sizes of the degenerate multiplets.

For the above reasons, several groups have been looking into
alternative ways to conserve symmetries within the DMRG algorithm.
In this section, we briefly review the important progress that has
been made.

The first significant progress along these lines was the work of
Sierra and Nishino \cite{IRF1}. In their work, they showed that
Hamiltonians with a continuous symmetry could be recast into the
form of Interaction Round a Face (IRF) Hamiltonians, familiar in
statistical mechanics. They then showed how to develop a DMRG
algorithm appropriate to IRF Hamiltonians. The method was applied
to several spin chains with impressive success. Examples include
the spin-1/2 \cite{IRF1}, spin-1 and spin-2 \cite{IRF2} Heisenberg
chains and the XXZ chain \cite{IRF1}. In all cases considered, the
method achieved significantly greater accuracy than the usual DMRG
algorithm when keeping the same number of states.

Despite the historical importance of the IRF-DMRG algorithm in the
proper incorporation of symmetries in the DMRG, it is difficult to
imagine its use for more complex systems with more complicated
symmetries.  Where major progress was made towards the development
of a general framework for incorporating symmetries in the DMRG
was through the work of McCulloch and Gul\'{a}csi \cite{MG1}.
Their work presented a formalism that could be applied to any
group of symmetry transformations, through proper construction of
basis states that preserve the symmetry. They refer to this as the
non-Abelian DMRG method.  In the following subsection, we describe
in some detail how their method can be applied to a finite Fermi
system with conserved total angular momentum. We will refer to
this specific example of their method as the jDMRG algorithm.

\subsection{Details on the jDMRG algorithm}

In htis subsection, we develop the jDMRG formalism, showing in
some detail how it can be implemented for finite Fermi systems so
as to produce exact angular momentum eigenstates. For simplicity,
we restrict the discussion to systems with but one type of
particle. For systems with two or more types of particles, atomic
nuclei being an example, some slight modifications are required.

We begin by stating the problem. We wish to solve the
Schr\"{o}dinger equation for a set of $N$ identical particles
subject to a Hamiltonian written in second-quantized form as

\begin{equation}
H=-\sum_{j}\widehat{j}\ \varepsilon _{j}\left( a _{j}^{\dagger
}\widetilde{a }_{j}\right)
_{0}^{0}+\sum_{j_{1}j_{2}j_{3}j_{4}L}X_{j_{1}j_{2}j_{3}j_{4}}^{L}\left[
\left( a _{j_{1}}^{\dagger }a _{j_{2}}^{\dagger }\right)
^{L}\left( \widetilde{a}_{j_{3}}\widetilde{a }_{j_{4}}\right)
^{L}\right] _{0}^{0} ~,
\end{equation}
\noindent where $\widehat{j}=\sqrt{2j+1}$.

Note that we are assuming here that the particles under discussion
are
created (annihilated) in the single-particle state $\{jm\}$ by the operator $%
a^{\dagger}_{jm}~~ (a_{jm})$ and that the Irreducible Tensor
Operator
associated with the single-particle annihilation operator is defined by $%
\widetilde{a}_{jm}=\left( -\right) ^{j-m}a_{j-m}$. By working in
terms of the Irreducible Tensor Operators $a^{\dagger}_{jm}$ and
$\widetilde{a}_{jm}$, we are able to couple operators directly and
thus properly reflect the scalar nature of the Hamiltonian. Note
that we use a coupling notation whereby $(...)^J_M$ and
$[...]^J_M$ refer to coupled Irreducible Tensor Operators of rank
$J$ and projection $M$.

The $N$ particles will be restricted to a finite set of
single-particle levels $\{jm\}$, as depicted schematically in
figure \ref{jDMRG}a for five orbits. The orbits are shown in
ascending order according to their single-particle energies
$\varepsilon_j$.

The jDMRG procedure involves three stages. The first is a
preliminary stage, in which all necessary information on the
individual active single-particle orbits are calculated and
stored. The second is the ``warm-up" stage, familiar to all DMRG
algorithms. And the third is the likewise familiar ``sweeping
stage".

\subsubsection{The preliminary stage.}

As the procedure will include the effects of all active
single-particle levels in an angular-momentum coupled
representation, we must store the following:

\begin{itemize}
\item  All single-shell states $|j_i,n_i, \alpha_i, J_i>$, for the various
active orbits $\{j_i\}$. Here $n_i$ refers to the number of
particles in the state and $\alpha$ refers to any additional
quantum numbers needed to fully describe the state of total
angular momentum $J_i$.

\item  The one-particle Coefficients of Fractional Parentage (CFPs) for all active
orbits. These are the expansion coefficients required to express
an antisymmetrized $n$-particle state as a product of
antisymmetrized states of $n-1$ particles and a single particle.

\item  The reduced (or double-bar) matrix elements of all possibly relevant operators for
those orbits, namely $a^{\dagger}_j$, $[a^{\dagger}_j a^{\dagger}_j ]^K$, $%
[a^{\dagger}_j \tilde{a}_j ]^K$, $[\{a^{\dagger}_j a^{\dagger}_j\}\tilde{a}%
_j ]^K$ and $[\{a^{\dagger}_j a^{\dagger}_j\}^K\{\tilde{a}_j
\tilde{a}_j \}^K]^0$.
\end{itemize}

The reduced matrix elements of $\tilde{a}_j$ and $a^{\dag}_j$ are
very simply related to the one-particle CFPs. The reduced matrix
elements of the more-complicated product operators can be obtained
from them using well-known formulae \cite{DST}.

\subsubsection{The warm-up stage.}

\begin{figure}
\begin{center}
\hspace{1.cm}\epsfxsize=7cm\epsfysize=6cm\epsfbox{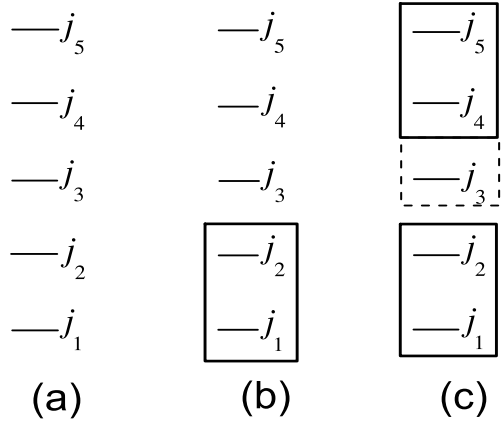}
\end{center}
\caption{\label{jDMRG}Schematic representation of the steps of the
the jDMRG method.}
\end{figure}

We now discuss the warm-up phase, in which we build a first
approximation to the dominant structure within groups of active
orbits. As we will see, most of the formalism we develop for the
warm-up phase will also be needed in the sweep phase. In fig.
\ref{jDMRG}b, we schematically illustrate the ``warm-up" phase for
the same set of five orbits.

We start off with the first orbit $j_1$, for which all the
single-shell matrix elements were calculated in the ``preliminary"
stage. We then add the second orbit, forming an enlarged ``block".
This is schematically illustrated in fig. \ref{jDMRG}b, where the
enlarged block contains orbits $j_1$ and $j_2$.

Next, we must build the reduced matrix elements of all relevant
sub-operators of the Hamiltonian for the enlarged block. To see
how this is done, we consider a  generic block involving $k$
orbits, $\left| j_{1}\cdots
j_{k},n,\alpha ,J\right\rangle$. The enlarged block basis, in which orbit $%
j_{k+1}$ has been added, is $\left| j_{1}\cdots j_{k},n,\alpha
,K;j_{k+1},m,\beta, L ~(J)\right\rangle$. The matrix elements of
the single-particle creation operator $a_i^{\dagger}$ in the
enlarged block are

\[
\left\langle j_{1}... j_{k},n^{\prime },\alpha ^{\prime
},K^{\prime };j_{k+1},m^{\prime },\beta ^{\prime },L^{\prime
}~(J^{\prime })\right\| a _{i}^{\dagger }\left\| j_{1}...
j_{k},n,\alpha ,K;j_{k+1},m,\beta, L~(J)\right\rangle
\]

\[
~~= \delta _{\beta \beta ^{\prime }}\delta _{mm^{\prime }}\delta
_{LL^{\prime }}\left( -\right) ^{K^{\prime }+L^{\prime }+J+i}~\widehat{J}%
\widehat{J}^{\prime }\left\{
\begin{array}{ccc}
K^{\prime } & J^{\prime } & L^{\prime } \\
J & K & i
\end{array}
\right\}
\]
\[
 ~~~~~~~~\times~ \left\langle j_{1}\cdots j_{k},n^{\prime },\alpha ^{\prime
},K^{\prime }\right\| a _{i}^{\dagger }\left\| j_{1}\cdots
j_{k},n,\alpha ,K\right\rangle +
\]

\[
~~~~~~~~+\delta _{\alpha \alpha ^{\prime }}\delta _{nn^{\prime
}}\delta
_{KK^{\prime }}\left( -\right) ^{K^{\prime }+L+J^{\prime }+i}~\widehat{J}%
\widehat{J}^{\prime }\left\{
\begin{array}{ccc}
L^{\prime } & J^{\prime } & K^{\prime } \\
J & L & i
\end{array}
\right\} \]
\begin{equation}
~~~~~~~~\times~\left\langle j_{k+1},m^{\prime },\beta ^{\prime
},L^{\prime }\right\| a _{i}^{\dagger }\left\| j_{k+1},m,\beta,
L\right\rangle  ~.\label{coupledmes}
\end{equation}

It involves a reduced matrix element from the previous iteration
(which has already been calculated) and a reduced matrix element
for the orbit that is being added (which was calculated in the
preliminary stage). Thus, all information is available to
calculate this reduced matrix element, and indeed all others, in
the enlarged block.

As in all DMRG algorithms, we now wish to choose out of all the
states in the enlarged block those to keep. Here the strategy is
somewhat different depending on whether there is only one kind of
particle or two. If there are two kinds of particle, one can
couple the corresponding states together, build and diagonalize
the superblock Hamiltonian and then truncate in the sector of
interest by using its reduced density matrix, assuming each type
of particle is the medium for the other.

When there is only one kind of particle, as in the example under
discussion, an alternative strategy must be found. A natural one
is to use the Wilson RG strategy of diagonalizing the Hamiltonian
in the enlarged space and keeping the $m$ lowest eigenstates for
each value of $n$ and $J$.

Following truncation, the reduced matrix elements for the
truncated enlarged block are calculated and stored on hard disk.
This is done for all size blocks, $j_1$, $j_1 \rightarrow j_2$,
$j_1 \rightarrow j_3$, $j_1 \rightarrow j_4$, and $j_1 \rightarrow
j_5$.

\subsubsection{The sweep stage.}

We now discuss the sweep stage, whereby we sweep back and forth
through the single-particle levels to improve the description of
each block by optimally taking into account its correlation with
{\em all} other levels.

The sweep stage is schematically illustrated in fig. \ref{jDMRG}c
for the same five orbits. At the point represented by the figure,
we are sweeping backwards through the single-particle levels and
have treated as a block the last two levels, $j_5$ and $j_4$. We
wish to enlarge this block to include level $j_3$. The medium for
this enlarged will be levels $j_1 \rightarrow j_2$.

The sweep stage involves first a calculation of all relevant
matrix elements in the enlarged block consisting of levels $j_3
\rightarrow j_5$. This can be done using exactly the same formulae
developed for the warm-up stage, with one example given in Eq.
(\ref{coupledmes}).

Once we have calculated all matrix elements in the enlarged block,
we construct the states of the superblock by coupling those of the
enlarged block (orbits $j_3 \rightarrow j_5$) to those of the
medium (orbits $j_1 \rightarrow j_2$), {\em viz:}  $\left|
j_{1}\cdots j_{2},n,\alpha ,K;j_{3}\cdots j_5 ,m,\beta, L
~(J)\right\rangle$. We only construct states with the correct
total number of particles $N=n+m$.

The calculation of the Hamiltonian matrix in the superblock
involves a sum of terms, each being a product of a reduced matrix
element in the enlarged block (just calculated) and a reduced
matrix element in the medium (output from the previous sweep or
the warm up).

As always, the sweep stage is iterated until acceptable
convergence has been achieved.

\subsection{Applications of the non-Abelian DMRG formalism}

The first application \cite{MG1} of the non-Abelian DMRG method
was reported by McCulloch and Gul\'{a}csi for the one-dimensional
Hubbard model. This model has the usual spin symmetry and also a
pseudo-spin symmetry generated by the set of operators $I^+=\sum_i
(-)^i c^{\dagger}_{i\uparrow} c^{\dagger}_{i\downarrow}$,
$I^-=\sum_i (-)^i c_{i\downarrow} c_{i\uparrow}$ , and
$I^z=\sum_i\frac{1}{2}(n_{i\uparrow} + n_{i\downarrow}-1)$. Thus
the DMRG algorithm can be implemented at three levels: (1) by only
including the spin projection as a conserved quantum number
(called $U(1)$), (2) by including the total spin as a conserved
quantum number (called $SU(2)$) and (3) by considering both the
total spin and the pseudo-spin as good quantum numbers (called
$SO(4)$).  There is also an additive (or Abelian) symmetry of
number conservation (likewise $U(1)$), which can be included at
each of the three levels. Results for the three levels of DMRG
application are summarized in table \ref{tableMG1} for a
half-filled 60-site Hubbard lattice with $t=U=1$.

\begin{table}
\caption{\label{tableMG1}Comparison of results for three sets of
DMRG calculations for the ground state of the half-filled 60-site
Hubbard model with $t=U=1$. The CPU time is in seconds per sweep.
(1) $U(1)\times U(1)$ refers to a calculation in which number
conservation and conservation of spin projection were imposed; (2)
$U(1) \times SU(2)$ also includes total spin conservation; (3)
$SO(4)$ in addition includes pseudo-spin conservation. Extracted
from Table I of ref. \protect\cite{MG1}.   }
\begin{indented}
\item[]\begin{tabular}{cccccr} \br
Basis &  $m$ & $D$ &  $E$  & $(E-E_g)/|E_g|$ & CPU \\
\mr
$U(1)\times U(1)$  & $100$ & $100$ &   $-76.7484986435$ &  $4.2\times10^{-5}$ & $10$ \\
$U(1) \times U(1)$ & $300$ & $300$ & $-76.7516910404$ & $6.3
\times10^{-7}$ & $110$ \\
$U(1) \times SU(2)$ & $100$ & $226$ & $-76.7515581914$ & $2.3
\times
10^{-6}$ & $15$ \\
$U(1) \times SU(2)$ & $300$ & $716$ & $-76.7517389831$ & $1.1
\times
10^{-8}$ & $158$ \\
$SO(4)$ & $100$ & $526$ & $-76.7517351742$ & $6.1 \times
10^{-8}$ & $18$ \\
$SO(4)$ & $300$ & $1766$ & $-76.7517398448$ & $7.9 \times
10^{-11}$ & $133$ \\
\br

\end{tabular}
\end{indented}
\end{table}

The table includes a quantity $D$, which in the case of a
non-Abelian symmetry is equal to the number of states in the
Abelian analysis required to get the same results as in the
symmetry-adapted analysis. Thus, for example, a calculation in
which the full SO(4) symmetry is preserved and in which a total of
m=100 states are maintained in a block would require $m=526$
states for comparable accuracy when only the z-projection and the
particle number are conserved. The table also indicates the
computational effort required for these various calculations.
Keeping the same number of states in a calculation with full
symmetry restoration as in a calculation with only additive
symmetries increases the cost in CPU time very little, but yields
an enormous improvement in accuracy.

The results shown in Table \ref{tableMG1} are for the the S=0
ground state of the system. Calculations of higher spin states do
not lead to as dramatic an improvement and cost savings.

\begin{table}
\caption{\label{tableMG2}Comparison of results of DMRG
calculations for the ground state of a 16$\times$6 t-J system with
$J=0.35$, $t=1$, eight holes and cylindrical boundary conditions.
The results labelled U(1) only preserve spin projection and are
from ref. \protect\cite{WS}. Those labelled SU(2) involve full
spin conservation. An extrapolated ``true" ground state energy is
also shown. From Table I of ref. \protect\cite{MG2}. Reprinted by
permission of Taylor \& Francis Ltd.,
http://www.tandf.co.uk/journals/titles/14786435.}
\begin{indented}

\item[]\begin{tabular}{ccc}
\br
Basis &  $m$ & $E$  \\
\mr
$U(1)$  & $1000$ &    $-52.279$  \\
$SU(2)$ & $500$ & $-52.284$ \\
$SU(2)$ & $800$ & $-52.463$  \\
$SU(2)$ & $1200$ & $-52.520$ \\
$-$ & $\infty$ & $-52.65 \pm 0.05$ \\
\br
\end{tabular}
\end{indented}
\end{table}

Subsequently, the method was applied to the two-dimensional t-J
Model for lattices up to 24$\times$6 \cite{MG2}. Here the purpose
was no longer to simply assess the usefulness of symmetry
restoration in the DMRG approach, but rather to use the power of
the method to provide useful insight into this model of possible
relevance to high-$T_c$ superconductors. In particular, it
attempted to address the important, but controversial, issue of
stripe formation in the t-J model.

The analysis made use of the Liang and Pang approach to 2D
lattices \cite{LiangPang}, which unrolls the 2D lattice onto a 1D
lattice with long-range interactions. The calculations included up
to $m=1200$ states per block. An indication of the accuracy of the
method is given in Table \ref{tableMG2}, where the results
obtained for a 16$\times$6 lattice with full spin conservation are
compared with those having conserved spin projection only
\cite{WS}. As  can be readily seen, the calculations with full
SU(2) spin conservation lead to significantly lower energies for
comparable block sizes. The ground state emerges from this
calculation in a two-stripe configuration, in agreement with ref.
\cite{WS}.

More recently, the method was applied to the problem of localized
spin ordering in Kondo lattice models \cite{MG3}. By exactly
preserving spin symmetry in the problem, the authors were able to
directly measure the magnetization of the ground state and this
made it possible for them to confirm the existence of a second
ferromagnetic phase in the model \cite{TSU}, for which weak
signals had been observed in a previous DMRG study that did not
fully restore symmetries \cite{SUNI}.

\section{\protect\bigskip Summary and Outlook}

In this review, we have described the current status of efforts to
develop the Density Matrix Renormalization Group for application
to finite Fermi system. Enormous progress has been made on this
front and the method has indeed been applied with impressive
success in several areas, especially in applications to Quantum
Chemistry, to study the physics of ultrasmall superconducting
grains and to study the properties of two-dimensional electron
systems. Along with these applications has come much work aimed at
identifying the key issues that must be resolved in order to build
an optimized DMRG algorithm for use in these and other systems.
One of the most important issues concerns the optimal approach to
ordering the single-particle basis states that are iteratively
added by the DMRG algorithm. While important progress has been
made, questions still remain and more work remains to be done.

Another area in which important progress has been made concerns
the full incorporation of symmetries in the DMRG method. It is now
possible to consistently incorporate non-Abelian symmetries, such
as rotational invariance, in addition to additive or Abelian
symmetries. Early applications of the associated new methodology
suggest that it permits increased accuracy in DMRG calculations,
while requiring less CPU cost. More work along these lines is
clearly warranted.

Another interesting point that has recently emerged concerns the
usefulness of the DMRG method when treating weakly-coupled
strongly-correlated subsystems. An application of the
particle-hole DMRG method to tunnelling between ultrasmall
superconducting grains suggests that the methodology breaks down
when the tunnel coupling between the two grains is very weak.
Similar issues also arise when dealing with a set of
single-particle levels that involve well-separated subsets ({\em
e.g.} shells), which are also very weakly coupled \cite{PD}. It
would be interesting to see whether the DMRG algorithm can be
appropriately modified to handle such scenarios. A possibility
might be to apply the DMRG separately to the various subsystems
and then to couple them perturbatively.

The issue of the ordering of levels for their optimal treatment
within the DMRG method has recently led to the suggestion that
Quantum Information Theory might provide a useful handle on the
problem. This connection needs further exploration.

We have focussed our remarks on some of the most critical
questions that have recently been raised in the context of the
DMRG method as it applies to finite Fermi systems and some of the
most interesting methodological breakthroughs. As further progress
is made to address these and other issues regarding the optimal
protocol for DMRG studies, we expect a rapid increase in the
number of applications. A potentially fertile area that has not
yet been explored is the atomic nucleus.

In closing, we believe that the Density Matrix Renormalization
Group method will continue to grow as a viable method for carrying
out reliable and extremely accurate calculations for the
properties of complex finite Fermi systems, adding to the enormous
history of success it has already achieved in the description of
quantum lattices.

\ack We would like to express our appreciation to Mario Stoitsov,
Sevdalina Dimitrova, German Sierra, Nicu Sandulescu and Larisa
Pacarescu, all of whom made important contributions to the DMRG
projects on which we have worked. Helpful discussions with Willy
Dussel, David Dean and especially Thomas Papenbrock on aspects of
the DMRG method are also gratefully acknowledged. This work was
supported in part by the Spanish DGI under grant
BFM2003-05316-C02-02 and in part by the US National Science
Foundation under grant \#s PHY-9970749 and PHY-0140036.

\end{document}